\documentclass[twocolumn]{aastex62}

\usepackage{mathtools}
\usepackage{amsmath}
\usepackage{subfigure}
\usepackage{color}
\usepackage{fnpct}
\usepackage[multiple]{footmisc}
\usepackage{enumitem}
\usepackage{autobreak}
\usepackage{breakurl}
\usepackage{graphicx} 
\usepackage{multirow}

\setenumerate[1]{itemsep=0pt,partopsep=0pt,parsep=\parskip,topsep=5pt}
\setitemize[1]{itemsep=0pt,partopsep=0pt,parsep=\parskip,topsep=5pt}
\setdescription{itemsep=0pt,partopsep=0pt,parsep=\parskip,topsep=5pt}

\newcommand{\black}[1]{\textcolor{black}{\textbf{\boldmath #1}}}

\received{---}
\revised{---}
\accepted{---}

\shorttitle{BASS DR3}
\shortauthors{Hu Zou et al.}

\begin{document}

\title{The Third Data Release of the Beijing-Arizona Sky Survey}

\correspondingauthor{Hu Zou}
\email{zouhu@nao.cas.cn}

\author{Hu Zou}
\affil{Key Laboratory of Optical Astronomy, National Astronomical Observatories, Chinese Academy of Sciences, Beijing 100012, China }
\author{Xu Zhou}
\affil{Key Laboratory of Optical Astronomy, National Astronomical Observatories, Chinese Academy of Sciences, Beijing 100012, China }
\author{Xiaohui Fan}
\affil{Steward Observatory, University of Arizona, Tucson, AZ 85721}
\author{Tianmeng Zhang}
\affil{Key Laboratory of Optical Astronomy, National Astronomical Observatories, Chinese Academy of Sciences, Beijing 100012, China }
\author{Zhimin Zhou}
\affil{Key Laboratory of Optical Astronomy, National Astronomical Observatories, Chinese Academy of Sciences, Beijing 100012, China }
\author{Xiyan Peng}
\affil{Key Laboratory of Optical Astronomy, National Astronomical Observatories, Chinese Academy of Sciences, Beijing 100012, China }
\author{Jundan Nie}
\affil{Key Laboratory of Optical Astronomy, National Astronomical Observatories, Chinese Academy of Sciences, Beijing 100012, China }
\author{Linhua Jiang}
\affil{Kavli Institute for Astronomy and Astrophysics, Peking University, Beijing 100871, China}
\author{Ian McGreer}
\affil{Steward Observatory, University of Arizona, Tucson, AZ 85721}
\author{Zheng Cai}
\affil{Department of Physics and Tsinghua Center for Astrophysics, Tsinghua University, Beijing 100086, China}
\author{Guangwen Chen}
\affil{Key Laboratory for Research in Galaxies and Cosmology, Department of Astronomy, University of Science and Technology of China, Hefei 230026, China}
\author{Xinkai Chen}
\affil{Key Laboratory for Research in Galaxies and Cosmology, Department of Astronomy, University of Science and Technology of China, Hefei 230026, China}
\author{Arjun Dey}
\affil{National Optical Astronomy Observatory, Tucson, AZ 85719}
\author{Dongwei Fan}
\affil{Key Laboratory of Optical Astronomy, National Astronomical Observatories, Chinese Academy of Sciences, Beijing 100012, China }
\author{Joseph R. Findlay}
\affil{Department of Physics and Astronomy, University of Wyoming, Laramie, WY 82071, USA}
\author{Jinghua Gao}
\affil{Key Laboratory of Optical Astronomy, National Astronomical Observatories, Chinese Academy of Sciences, Beijing 100012, China }
\author{Yizhou Gu}
\affil{Department of Physics, Nanjing Normal University, WenYuan Road 1, Nanjing 210046, China}
\author{Yucheng Guo}
\affil{Kavli Institute for Astronomy and Astrophysics, Peking University, Beijing 100871, China}
\author{Boliang He}
\affil{Key Laboratory of Optical Astronomy, National Astronomical Observatories, Chinese Academy of Sciences, Beijing 100012, China }
\author{Zhaoji Jiang}
\affil{Key Laboratory of Optical Astronomy, National Astronomical Observatories, Chinese Academy of Sciences, Beijing 100012, China }
\author{Junjie Jin}
\affil{Key Laboratory of Optical Astronomy, National Astronomical Observatories, Chinese Academy of Sciences, Beijing 100012, China }
\author{Xu Kong}
\affil{Key Laboratory for Research in Galaxies and Cosmology, Department of Astronomy, University of Science and Technology of China, Hefei 230026, China}
\author{Dustin Lang}
\affil{David Dunlap Institute, University of Toronto, Toronto, Canada}
\author{Fengjie Lei}
\affil{Key Laboratory of Optical Astronomy, National Astronomical Observatories, Chinese Academy of Sciences, Beijing 100012, China }
\author{Michael Lesser}
\affil{Steward Observatory, University of Arizona, Tucson, AZ 85721}
\author{Feng Li}
\affil{Department of Physics, Nanjing Normal University, WenYuan Road 1, Nanjing 210046, China}
\author{Zefeng Li}
\affil{Kavli Institute for Astronomy and Astrophysics, Peking University, Beijing 100871, China}
\author{Zesen Lin}
\affil{Key Laboratory for Research in Galaxies and Cosmology, Department of Astronomy, University of Science and Technology of China, Hefei 230026, China}
\author{Jun Ma}
\affil{Key Laboratory of Optical Astronomy, National Astronomical Observatories, Chinese Academy of Sciences, Beijing 100012, China }
\author{Moe Maxwell}
\affil{Steward Observatory, University of Arizona, Tucson, AZ 85721}
\author{Xiaolei Meng}
\affil{Department of Physics and Tsinghua Center for Astrophysics, Tsinghua University, Beijing 100086, China}
\author{Adam D. Myers}
\affil{Department of Physics and Astronomy, University of Wyoming, Laramie, WY 82071, USA}
\author{Yuanhang Ning}
\affil{Kavli Institute for Astronomy and Astrophysics, Peking University, Beijing 100871, China}
\author{David Schlegel}
\affil{Lawrence Berkeley National Labortatory, Berkeley, CA 94720}
\author{Yali Shao}
\affil{Kavli Institute for Astronomy and Astrophysics, Peking University, Beijing 100871, China}
\author{Dongdong Shi}
\affil{Steward Observatory, University of Arizona, Tucson, AZ 85721}
\author{Fengwu Sun}
\affil{Kavli Institute for Astronomy and Astrophysics, Peking University, Beijing 100871, China}
\affil{Steward Observatory, University of Arizona, Tucson, AZ 85721}
\author{Jiali Wang}
\affil{Key Laboratory of Optical Astronomy, National Astronomical Observatories, Chinese Academy of Sciences, Beijing 100012, China }
\author{Shu Wang}
\affil{Kavli Institute for Astronomy and Astrophysics, Peking University, Beijing 100871, China}
\author{Yonghao Wang}
\affil{Key Laboratory of Optical Astronomy, National Astronomical Observatories, Chinese Academy of Sciences, Beijing 100012, China }
\author{Peng Wei}
\affil{Key Laboratory of Optical Astronomy, National Astronomical Observatories, Chinese Academy of Sciences, Beijing 100012, China }
\affil{Xinjiang Astronomical Observatory, Chinese Academy of Sciences, Urumqi 830011, China}
\author{Hong Wu}
\affil{Key Laboratory of Optical Astronomy, National Astronomical Observatories, Chinese Academy of Sciences, Beijing 100012, China }
\author{Jin Wu}
\affil{Kavli Institute for Astronomy and Astrophysics, Peking University, Beijing 100871, China}
\author{Xiaohan Wu}
\affil{Kavli Institute for Astronomy and Astrophysics, Peking University, Beijing 100871, China}
\affil{Harvard-Smithsonian Center for Astrophysics, 60 Garden Street, Cambridge, MA 02138, USA}
\author{Jinyi Yang}
\affil{Steward Observatory, University of Arizona, Tucson, AZ 85721}
\author{Qian Yang}
\affil{Kavli Institute for Astronomy and Astrophysics, Peking University, Beijing 100871, China}
\author{Qirong Yuan}
\affil{Department of Physics, Nanjing Normal University, WenYuan Road 1, Nanjing 210046, China}
\author{Minghao Yue}
\affil{Kavli Institute for Astronomy and Astrophysics, Peking University, Beijing 100871, China}
\affil{Steward Observatory, University of Arizona, Tucson, AZ 85721}



\begin{abstract}
The Beijing-Arizona Sky Survey (BASS) is a wide and deep imaging survey that covers a 5400 deg$^2$ area in the northern Galactic cap with the 2.3m Bok telescope using two filters ($g$ and $r$ bands). The Mosaic $z$-band Legacy Survey (MzLS) covers the same area in $z$ band with the 4m Mayall telescope. These two surveys will be used for spectroscopic targeting by the Dark Energy Spectroscopic Instrument (DESI) in the most northern portion of the DESI footprint. This paper describes the third data release (DR3) of BASS, which contains the photometric data from all BASS and MzLS observations between January, 2015 until the completion of BASS in March, 2019. The median astrometric precision relative to {\it Gaia} positions is about 17\,mas and the median photometric offset relative to the PanSTARRS1 photometry is within 5\,mmag. The median $5\sigma$ AB magnitude depths for point sources are 24.2, 23.6, and 23.0 mag for $g$, $r$, and $z$ bands, respectively. The photometric depth within the survey area is highly homogeneous, and the difference between the  20\% and 80\% depth is less than 0.3\,mag. The DR3 data, including raw data, calibrated single-epoch images, single-epoch photometric catalogs, stacked images, and co-added photometric catalogs, are publicly accessible at \url{http://batc.bao.ac.cn/BASS/doku.php?id=datarelease:home}.
\end{abstract}



\section{Introduction} 
The Beijing-Arizona Sky Survey (BASS\footnote{\url{http://batc.bao.ac.cn/BASS/}}) is a $g$ and $r$-band imaging survey in the northern Galactic cap using Steward Observatory's 2.3m Bok telescope on Kitt Peak, Arizona \citep{zou17b}. It is one of the three optical imaging surveys that aim at providing galaxy and quasar targets for follow-up by the Dark Energy Spectroscopic Instrument \citep[DESI;][]{des16}. The other two imaging surveys are the Dark Energy Camera Legacy Survey \citep[DECaLS;][]{blu16} and MOSAIC $z$-band Legacy Survey \citep[MzLS;][]{sil16}.  These three surveys jointly image more than 14,000 deg$^2$ of the sky at high Galactic latitude using telescopes at the Kitt Peak National Observatory and the Cerro Tololo Inter-American Observatory \citep[see][for a full summary of DECaLS and MzLS]{arjun19}. 

DESI is a next-generation dark energy experiment that will precisely measure the expansion rate and structure growth of the universe by obtaining spectroscopic redshifts of several tens of millions of galaxies and quasars \citep{des16}. In addition to providing spectroscopic targets for DESI, the three new imaging surveys noted above, together with the latest infrared data \citep{mei17,mei18} from the Wide-field Infrared Survey Explorer mission \citep[WISE;][]{wri10} will allow studies of Galactic structure, galaxy clusters, AGN evolution, high-redshift quasars, large-scale structure, and so on \citep[e.g.][]{zou17b}. Existing public data from the DESI imaging surveys have already been used to discover quasars at $z > 6$ \citep{wang17,wang18,yang19}, to detect ultra-diffuse dwarf galaxies \citep{shi17,zar19}, to search for changing-look AGNs and RR Lyrae Stars \citep{gar18,yang18}, to investigate the scaling relations of metal-poor star-forming galaxies \citep{gao18}, to estimate photometric redshifts \citep{zou19}, and to discover strongly lensed quasars \citep{shu18}.

BASS covers a sky area of about 5400 deg$^2$ in the northern Galactic cap in $g$ and $r$ bands. MzLS used the NOAO 4m Mayall telescope at Kitt Peak to cover almost the same area with $z$ band. The nominal imaging depths for these two surveys are $g = 24.0$, $r = 23.4$, and $z =22.5$ AB mag for 5$\sigma$ galaxies with an exponential light profile of half-light radius $r_\mathrm{h}=0\arcsec.45$ \citep{arjun19}. There were two previous public BASS data releases: Data Release 1 (DR1), which included only BASS data \citep{zou17a} and DR2, which included both BASS and MzLS data taken as of 2017 July \citep{zou18}. DR3 of BASS, which is presented in this paper, is the final data release. It includes all BASS observations from January, 2015 to March, 2019, together with all MzLS observations from November, 2015 to February, 2018. The goal of this paper is to summarize all aspects of these surveys, to describe how to access DR3, and to present data quality of this final release. 

The paper is organized as follows. Section \ref{sec-overview} provides an overview of the BASS and MzLS surveys. Section \ref{sec-reduction} describes the data reduction, including the image processing, astrometric calibration, and photometric calibration. The photometric pipeline is described in Section \ref{sec-photometry}. Section \ref{sec-quality} characterizes the data quality. Section \ref{sec-systemequ} provides equations for transforming the photometric systems of other commonly used surveys to the BASS/MzLS systems. Finally, Section \ref{sec-summary} summarizes the entirety of DR3.

\section{Overview of the BASS and MzLS surveys}  \label{sec-overview} 
DESI will conduct a large-scale spectroscopic survey using a 5000-fiber multi-object spectroscopic instrument installed on the Mayall 4-meter telescope at Kitt Peak National Observatory \citep{des16}. The DESI survey will measure spectroscopic redshifts for about 35 million galaxies and quasars over a five-year period between 2020 and 2025 in order to map the effect of dark energy on the expansion of the universe. The DESI spectroscopic targets will be selected using a combination of optical imaging data in $g$, $r$, and $z$ bands and near-infrared data in WISE $W1$ and $W2$ bands \citep{arjun19}.  DESI spectroscopic targets include bright galaxies (redshifts of $0.05 < z < 0.4$), luminous red galaxies ($0.4 < z < 1.0$), emission line galaxies ($0.6 < z < 1.6$), and quasars ($z <\sim 3.5$). Three new legacy optical imaging surveys were designed to provide the photometry needed for DESI targeting. BASS and MzLS are two of these three imaging surveys, and were designed to principally cover part of the northern Galactic cap \citep[see Figure 2 of][for the survey footprints]{zou17b}. 

BASS used the 2.3\,m Bok telescope to take $g$ and $r$-band imaging over a sky area of about 5400 deg$^2$ in the northern Galactic cap at $\delta > 30$\arcdeg. The Bok telescope is owned and operated by the University of Arizona's Steward Observatory. It is equipped with a wide-field prime-focus camera, 90Prime, with a field of view (FoV) of 1\arcdeg.08$\times$1\arcdeg.03 \citep{wil04}. The camera has a mosaic of four 4k$\times$4k CCDs, which were optimized for ultraviolet response. The peak quantum efficiency at 4000 {\AA} is close to 95\%. The CCD pixel scale is 0\arcsec.454 and the filling factor of the mosaic is about 94\%. 

MzLS used the 4\,m Mayall telescope to obtain $z$-band imaging over a similar sky area as BASS ($\delta > $ 32\arcdeg). The Mayall telescope is also located at Kitt Peak and is operated by the National Optical Astronomical Observatory (NOAO). The prime-focus camera, Mosaic-3, was populated with four 4k$\times$4k thick deep-depletion CCDs.  This improved $z$-band efficiency by 60\% compared with the previous Mosaic cameras. The camera FoV is about 36{\arcmin} and the CCD pixel scale is about 0\arcsec.26. 

Most DESI targets will be drawn from DECaLS imaging, which was obtained on a different instrument (DECam) at a different observatory (the Cerro Tololo Inter-American Observatory; CTIO). Therefore, achieving homogeneous target selections for DESI requires that the imaging survey strategy and photometric system are as consistent as possible between BASS, MzLS and DECaLS. The photometric systems of BASS and MzLS are close to the DECaLS system, which is the same as is used by the Dark Energy Survey \citep{des05}.   However, there are still some minor differences in the filter response curves, due to the difference in sites, filter manufacturing, optical systems, detector QE, etc. The central wavelengths at airmass of 1.0 are 4789, 6404, and 9210\,{\AA} for BASS $g$, $r$, and MzLS $z$ bands, respectively. For the BASS filter responses, we refer the reader to \citet{zou18} and the BASS webpage\footnote{\url{http://batc.bao.ac.cn/BASS/doku.php?id=datarelease:telescope\_and\_instrument:home\#filters}}. 

BASS was allocated a total of 387 nights from January, 2015 to March, 2019. MzLS was allocated 401 nights from October, 2015 to February, 2018. Both surveys shared a similar observing strategy: (1) the footprint was tiled by three dithered passes; (2) Pass 1 was observed in both photometric conditions and good seeing conditions ($< 1\arcsec.7$ for BASS and $< 1\arcsec.3$ for MzLS); (3) Pass 2 was observed in either photometric conditions or good seeing conditions; (4) Pass 3 was observed in all other conditions. The dithered passes were designed to fill the CCD gaps and ensure that at least one of the three passes was observed in photometric and good-seeing conditions. This strategy was implemented to benefit photometric calibration, the removal of particle events, and the detection of variable objects. In order to create a uniform survey, the exposure time was dynamically adjusted by an Exposure Time Calculator (ETC) according to the real-time observing conditions (e.g. sky brightness, transparency, and seeing). This approach ensured that different passes approached the same depth requirements. 

\begin{figure*}[!ht]
\centering
\includegraphics[width=\linewidth]{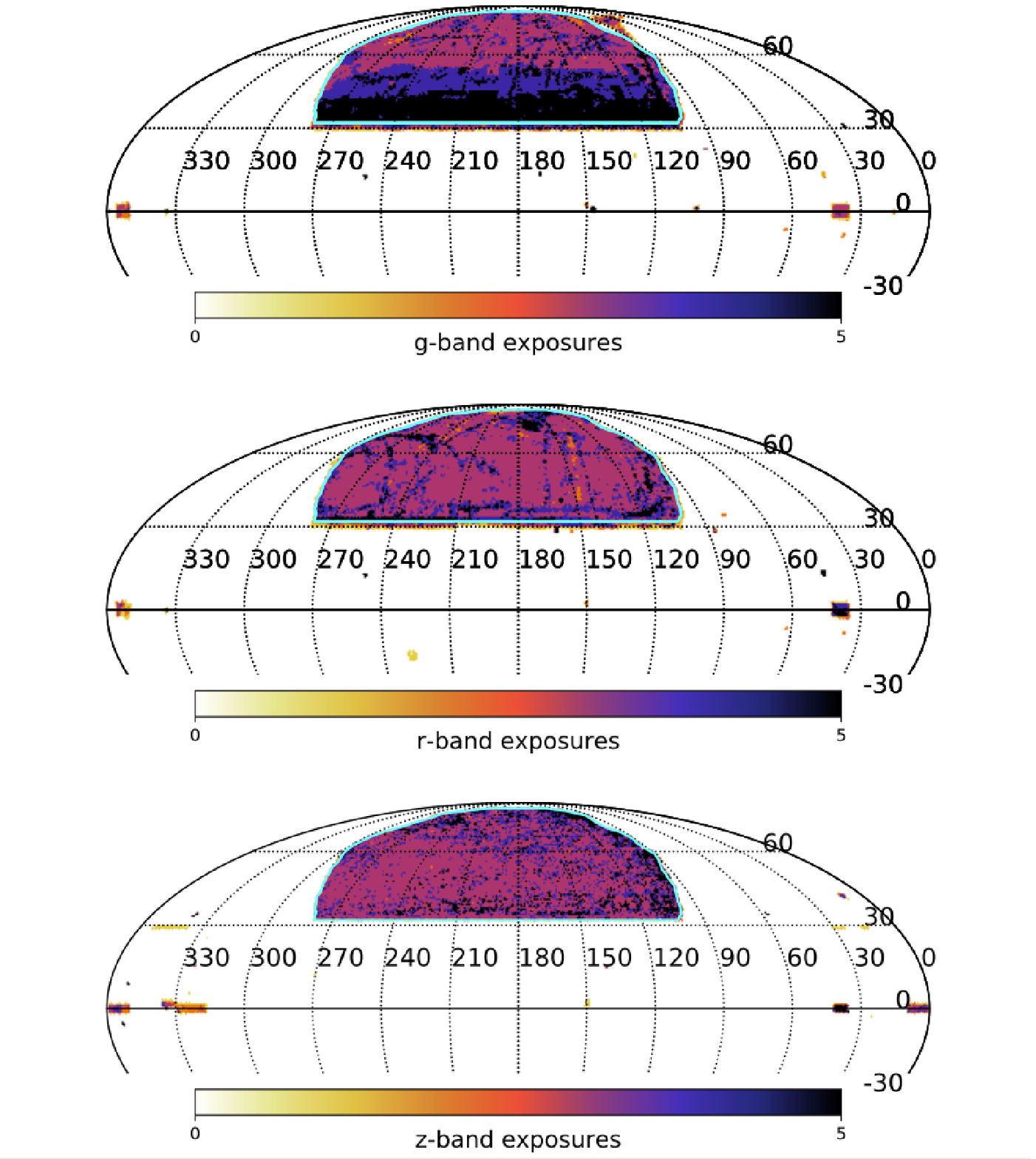}
\caption{Spatial distributions of exposure numbers for $g$, $r$, and $z$ bands. The currently expected DESI survey footprint is outlined in cyan.} \label{fig-cover} 
\end{figure*}

Figure \ref{fig-cover} shows the final coverage of all BASS and MzLS observations. In addition to the regular survey footprint as enclosed by the dashed cyan line in Figure \ref{fig-cover}, there are a few other patches across the sky, which are test regions (e.g. COSMOS\footnote{\url{http://cosmos.astro.caltech.edu/}}  and Stripe 82 of the Sloan Digital Sky Survey) and additional observations shared by other programs/observers. Most of the survey footprint is covered by three exposures for $r$ and $z$ bands. Those regions covered by more than three exposures correspond to regions that were re-observed because some of the three passes were initially of insufficient depth. Most notably, about one third of the $g$-band footprint is covered by more than three exposures. This is because $g$-band data taken in 2015, near BASS first light, was shallower due to issues with the ETC and additional noise from CCD readout amplifiers; the affected regions were later re-observed. Table \ref{tab-survey} summarizes some basic information about the BASS and MzLS surveys.  

\begin{table*}[!ht]
	\centering
	\small
	\caption{A summary of BASS and MzLS surveys}\label{tab-survey}
	  \begin{tabular}{r|cc}
	  \tableline\tableline
	  Telescope and Site & BASS & MzLS  \\
	  \tableline
	  Telescope & 2.3 m Bok & 4 m Mayall \\
	  Site & Kitt Peak & Kitt Peak \\
	  Elevation & 2071 m & 2071 m \\
	  \tableline
	  Camera & BASS & MzLS  \\
	  \tableline
	   Name & 90Prime & Mosaic-3 \\
	   CCD number & 4  & 4\\
	   CCD size & 4k$\times$4k & 4k$\times$4k \\
	   Gaps & 168{\arcsec} (RA), 54{\arcsec} (DEC) & 52{\arcsec} (RA), 62{\arcsec} (DEC) \\
	   Pixel scale & 0\arcsec.45 & 0\arcsec.26 \\
		FoV & 1\arcdeg.08$\times$1\arcdeg.03 & 36\arcmin$\times$36\arcmin \\
		Gain & $\sim$1.5 e/ADU & $\sim$1.8 e/ADU\\
		Readout noise & 8.4 e & 9 e\\
		Readout time & 35 s & 30 s\\
		Dark current & 0.8 e/hr & 0.9 e/hr \\ 
	   \tableline
	   Filter & BASS & MzLS  \\
	   \tableline
	   Name &  $g$ and $r$ &  $z$ \\ 
	   Effective wavelength & 4789 {\AA} and 6404 {\AA} & 9210 {\AA} \\
	   FWHM & 1435 {\AA}, 1420 {\AA} & 1430 {\AA} \\
	   \tableline
	   Survey parameters & BASS & MzLS  \\
	   \tableline
	   Area & 5400 deg$^2$ ($\delta > 30$\arcdeg) & 5100 deg$^2$ ($\delta > 32$\arcdeg) \\
	   Median seeing & $g=1\arcsec.6$, $r = 1\arcsec.4$   & $z=1\arcsec.0$ \\
	   Median depth\tablenotemark{a} & $g = 24.2$, $r=23.6$ & $z=23.0$ \\
	   Observation period & 2015.1--2019.3 & 2015.10--2018.2  \\
	  \tableline
	  \end{tabular}
	  \tablenotetext{a}{The depth is the magnitude limit for $5\sigma$ point sources from the combined photometry.}
\end{table*}

\section{Data Reduction} \label{sec-reduction}
BASS and MzLS operated on a principal of open access, with the raw data being released as soon as they were transferred to publicly accessible locations. The raw data were processed by dedicated data reduction pipelines to provide calibrated data products. These pipelines corrected for instrumental effects, and provided astrometric and photometric solutions. The calibrated data were then fed to a photometric pipeline to generate catalog products. 

Table \ref{tab-singlestats} provides information for all calibrated single-epoch CCD images, including the median seeing (in PSF FWHM), airmass, sky brightness, exposure time, and $5\sigma$ (PSF) magnitude limit. The median seeing is about 1\arcsec.6 for $g$ band, 1\arcsec.4 for $r$ band, and 1\arcsec.0 for $z$ band. The median $5\sigma$ depths for single-epoch images are 23.5, 23.0, and 22.5 mag for $g$, $r$, and $z$ bands, respectively.

\begin{table*}[!ht]
\centering
\caption{Median observational and quality statistics for calibrated single-epoch CCD images.}\label{tab-singlestats}
 \begin{tabular}{ccccccc}
 \tableline\tableline
 Filter & number of images &  airmass & exposure time & seeing & sky brightness & 5$\sigma$ depth  \\
 --        &   -- &   --       &        s       &     arcsec   &      mag$\cdot$arcsec$^{-2}$    & mag   \\
 \tableline
 $g$  & 101,351  & 1.08       &  103        &  1.58           &  22.18            & 23.50    \\
 $r$  & 79,489   & 1.10       &   102      &    1.43          &  21.13            & 22.99    \\
 $z$ &  243,121  & 1.11       &   90      &      1.02            &   18.78         & 22.45    \\
 \tableline
 \tableline
\end{tabular}

\end{table*}

\subsection{Imaging processing}
BASS and MzLS data were taken using different combinations of cameras and telescopes, which suffer from their own instrumental effects. The basic steps of the image processing are listed below. Additional details can be found in \citet{zou17a,zou18}. The listed operations were applied to both BASS and MzLS data, unless noted otherwise.
\begin{enumerate}
\item \black{Correction of line shifts (only in $z$ band and new in DR3).}
Occasionally, sub-blocks of an MzLS $z$-band image were shifted in one or more CCDs, resulting in the overscan being displayed in the data section of the image. The readout system can detect such events and marks them in the image header. These images were discarded since the data can not be recovered, except in one case where the data section of the image was entirely shifted by only one pixel to the overscan region. In this case the missing line was replaced with the corresponding line in the overscan region. 
\item \black{Cross-talk correction.}
Both BASS and MzLS data suffer from crosstalk in the readout systems. The intra-CCD and inter-CCD crosstalk coefficients were calculated using bright stars. These coefficients were recalculated annually, because the cameras were unloaded from the telescope and serviced every summer, which could potentially change the crosstalk characteristics. The crosstalk in the Mosaic-3 camera needed a larger correction than was required for 90Prime.
\item \black{Overscan and bias subtraction.}
The overscan correction was based on overscan lines in the CCD images, and a smoothed median overscan line was subtracted. For BASS data, the overscan might be contaminated by saturated stars. The contaminated regions were identified and masked during the overscan correction. The median bias was constructed from 20 bias frames taken before and after a night's observations. For BASS data taken in 2015, only the first bias frame was useful for bias subtraction due to the improper readout mode. A median bias was generated monthly and used to correct the data taken in each month.
\item \black{Flat-fielding.}
Both dome flats and super-sky flats were used for flat-fielding. The dome flat was used to correct pixel-to-pixel sensitivity variations. It was obtained from 20 flat frames taken with the telescope pointing at a white screen. The super-sky flat was used for illumination correction, which was constructed through combining all the science frames. 
\item \black{Amplifier gain balancing.}
The gain for each CCD amplifier varied with time, which could be noted in the sky background of the flat-fielded CCD image. We used the adjacent lines between four amplifiers of each CCD to balance the gain difference. 
\item \black{Removal of pattern noise (only $z$ band).}
The MzLS data suffered from serious pattern noise caused by electromagnetic interference during read-out. It would have effect on the background estimations and flux measurements of objects. The noise pattern for each CCD image is reconstructed through a low-pass filter. We apply a 5-order low-pass Butterworth digital filter\footnote{\url{https://docs.scipy.org/doc/scipy/reference/generated/scipy.signal.butter.html}} to the source-removed $z$-band image to create a smooth pattern. This pattern is subtracted from the original image. Figure \ref{fig-pattern} shows an example of removing the pattern noise. 
\begin{figure}[!ht]
\centering
\includegraphics[width=\linewidth]{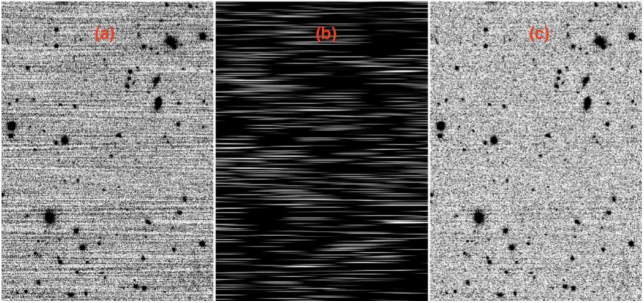}
\caption{Removal of $z$-band pattern noise. (a) partial raw image; (b) noise pattern constructed by a low-pass filter; (c) the image with the pattern subtracted.} \label{fig-pattern} 
\end{figure}
\item \black{Removal of pupil ghost (only $z$ band and new in DR3).} 
The pupil ghost appears in the center of the FoV of MzLS images and is quite obvious when the sky background is bright.  A ghost pattern image was created from the super-sky flat and the scaled pattern was subtracted from the CCD image to remove the pupil ghost. 
\item \black{Identifying CCD artifacts.}
Bad pixels were identified from the linearity measurements of flat frames taken with different exposure times. Other CCD artifacts were also identified, including saturated pixels, a black core in saturated stars due to integer overflow, bleeding trails from saturated stars, cosmic rays, and satellite tracks. 
\item \black{Interpolation over bad pixels and artifacts.}
Pixels that were affected by the CCD artifacts noted above were replaced by a bilinear interpolation of the surrounding, unaffected pixels.
\end{enumerate}

\begin{figure*}[!ht]
\centering
\includegraphics[width=0.8\linewidth]{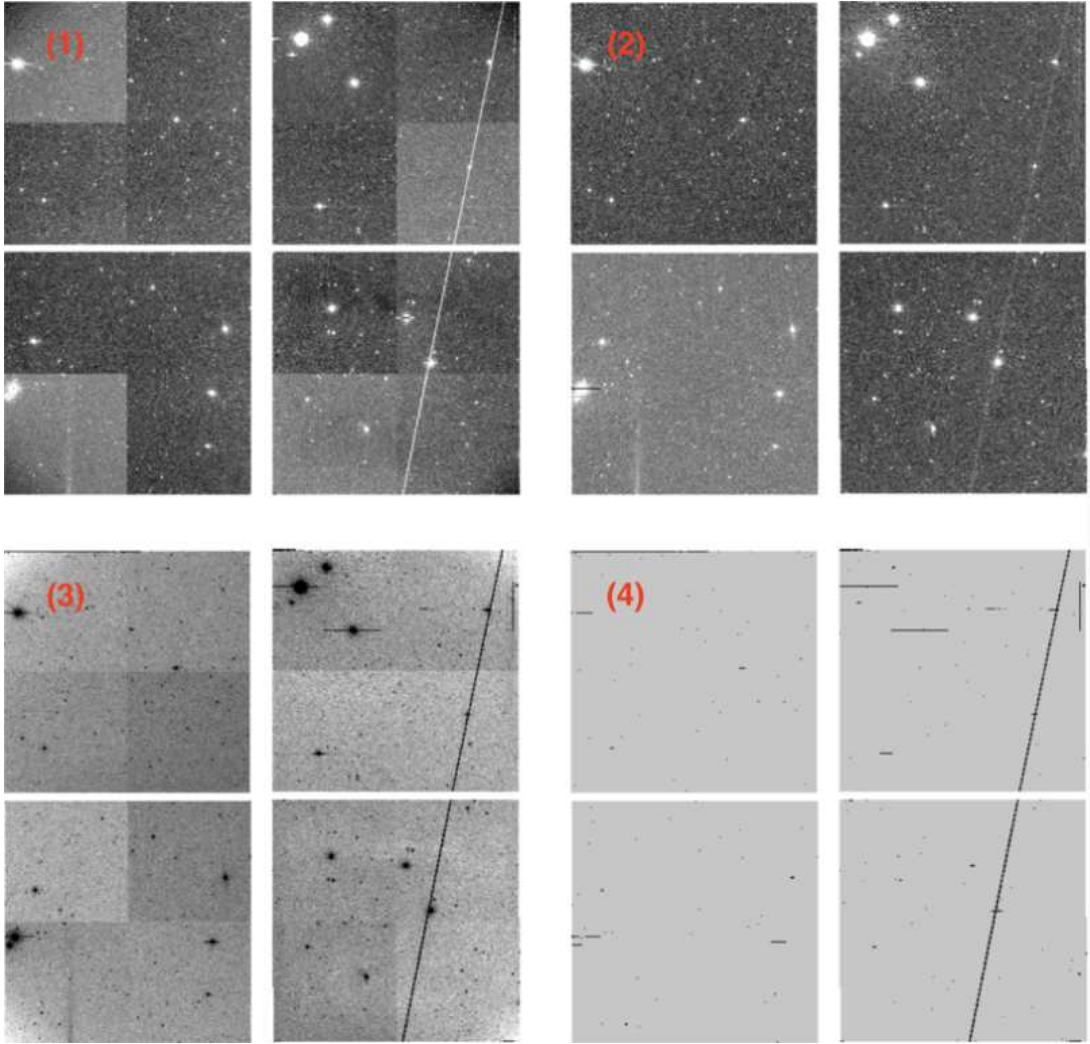}
\caption{An example of raw and processed data for one BASS $g$-band exposure. (1) raw images; (2) processed images; (3) weight images; (4) flag images. There are four images in each panel, representing the CCD mosaic. } \label{fig-image} 
\end{figure*}

Finally, three types of image maps were created, including detrended image, and weight, and flag maps. The weight map provides the inverse variance of flux for each pixel. The flag map tags each pixel with a flag to mark the type of bad pixels. The flag value is a sum of the power of 2 (a ``bit mask"): (1) 1 for bad pixel; (2) 2 for saturation; (3) 4 for cosmic ray; (4) 8 for satellite track. Figure \ref{fig-image} illustrates the raw and processed data. 

\subsection{Astrometric calibration} \label{sec-astrometry}
The astrometric calibration aims to derive a WCS solution that maps pixel coordinates to celestial coordinates. In addition to the standard WCS, a function was derived to characterize the focal distortion. The TPV world coordinate system was adopted for our astrometry. TPV builds on the standard TAN projection by adding a general polynomial distortion correction. We used the Software for Calibrating AstroMetry and Photometry \citep[SCAMP;][]{ber06} to calculate astrometric solutions. The SCAMP was run twice for each exposure in order to improve accuracy.

\begin{figure*}[!ht]
\centering
\includegraphics[width=0.9\textwidth]{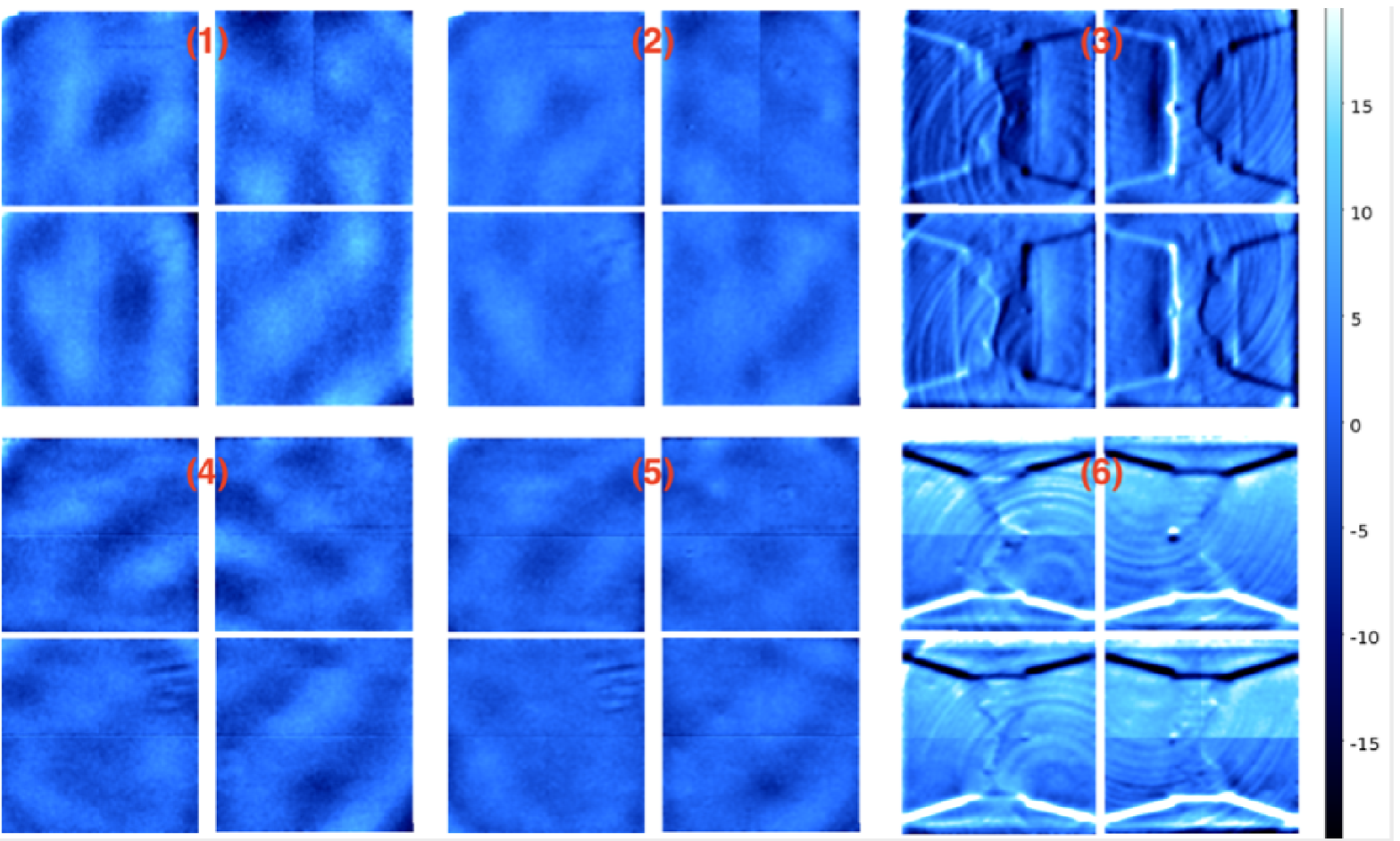}
\caption{Astrometric residuals in mas relative to {\it Gaia} DR2. (1)--(3) R.A. residuals for $g$, $r$, and $z$ bands, respectively; (4)--(6) decl. residuals for $g$, $r$, and $z$ bands, respectively. There are four images in each panel, representing the CCD mosaic.} \label{fig-astromres} 
\end{figure*}

Our calibration procedure effectively corrects for atmospheric refraction, at least for a mean color, and for focal distortions. Differential chromatic refraction is not considered in our astrometric solutions. 90Prime suffers from a charge transfer efficiency (CTE) effect, which affects centroid measurements. This effect results in a detectable astrometric offset at the amplifier boundary. The offset for CCD \#1 is biggest, comprising about 70 mas. Further investigations showed that the CTE effect produces linear offsets along the readout direction and that this effect is correlated with the brightness of sources (e.g.\ producing a larger gradient for fainter objects). We will have a dedicated paper about the astrometry for the BASS data, including the Bok distortion and 90Prime CTE effect (Peng et al. 2019; in preparation). Thus, before running SCAMP, we performed a correction of the CTE effect to the pixel positions according to the source brightnesses. The Mosaic-3 CCDs show residual astrometric patterns from bonding stresses and small-scale distortions known as ``tree rings" \citep{pla14}.  We constructed residual astrometric maps using coordinates of bright stars relative to the positions in the reference catalog, and corrected all possible astrometric residuals in our post-processed catalogs. The astrometric residual maps are shown in Figure \ref{fig-astromres}.

The reference catalogs for BASS DR1 were derived from the Sloan Digital Sky Survey (SDSS) and the Two Micron All Sky Survey \citep[2MASS;][]{skr06}. The median astrometric accuracy was measured to be about 150\,mas \citep{zou17a}. For BASS DR2, the {\it Gaia} DR1 catalog \citep{gai16a} was adopted. The {\it Gaia} data \citep{gai16b} provides superbly accurate astrometric measurements for objects with $G$ down to 20.5 over almost the entire sky. The overall astrometric accuracy for BASS DR2 was estimated to be 30--50\,mas \citep{zou18}. For DR3, we adopted the {\it Gaia} DR2 catalog \citep{gai18}, which is more complete in terms of sky coverage and has more accurate astrometry and photometry. Where available, the coordinates of {\it Gaia} sources are also corrected using {\it Gaia} proper motions.

\begin{figure*}[!ht]
\centering
\includegraphics[width=\linewidth]{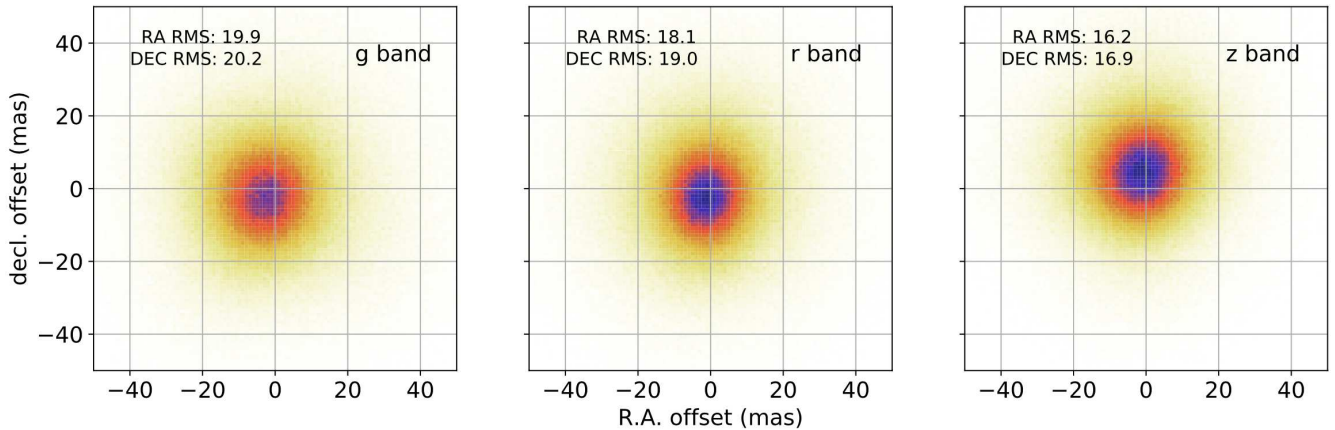}
\caption{Astrometric precisions relative to {\it Gaia} DR2 for single-epoch images in $g$, $r$, and $z$ bands. The color map shows the object density. The scatters of the astrometric offsets in R.A and decl. are displayed in each panel.} \label{fig-astrometry} 
\end{figure*}

Figure \ref{fig-astrometry} shows the accuracy of astrometric calibration in $grz$ bands for bright stars in single exposures relative to the positions in the {\it Gaia} DR2 catalog. The scatter in R.A. and decl. are about 19\,mas for BASS $g$ and $r$ bands and 17\,mas for MzLS $z$ band. 

\subsection{Photometric calibration} \label{sec-zeropoint}
The photometric calibration aims to derive a photometric zero point for each CCD image via comparing the observed instrumental magnitudes of stars with those from a reference catalog.

\citet{zho18} detailed our photometric calibration procedure, which is tied to the Pan-STARRS1 DR1 photometry \citep[hereafter PS1;][]{cha16} through a set of system transformation equations: 

\begin{equation}
\begin{split}
(g-i) \equiv &(g_{\rm PS1}-i_{\rm PS1}), \\
g_{\rm BASS} = &g_{\rm PS1} + 0.01077 + 0.10575(g-i) - \\
&0.02543(g-i)^2 + 0.00226(g-i)^3,  \\
r_{\rm BASS} = &r_{\rm PS1} + 0.00243 - 0.07650(g-i) + \\
&0.02809(g-i)^2 - 0.00967(g-i)^3, \\ 
z_{\rm MzLS} = &z_{\rm PS1} + 0.01029 - 0.08774(g-i) + \\
&0.03041(g-i)^2 - 0.00947(g-i)^3.
\end{split}
\end{equation}

The color terms were empirically derived by comparing BASS and PS1 photometry using stars with photometric errors $< 0.05$ mag and with color $0.4 < (g-i) < 2.7$. These constants were calibrated to the Pickles stellar spectral library \citep{pic98} using filter response curves. The above transformation equations were first applied to the PS1 photometry and then the zero points were calculated as the difference between the transformed PS1 magnitudes and the instrumental magnitudes. The instrumental magnitude was measured with an aperture of 26 pixels in diameter for both BASS and MzLS. Such an aperture is about $7\times$ PSF FWHM for median seeing conditions.

In addition to the external calibration described above, which ties the photometry to an external catalog, we also performed an internal calibration by using stars with repeated observations from different exposures. Our tiling strategy ensured that there were plenty of overlaps between different passes. The photometry for a specified exposure can be recalibrated by the stars in overlapping exposures, which provides a small correction to the external zero point. The internal calibration process we adopted was iterative, following these steps: (1) exposures with good photometric quality and external calibration were considered to be fiducial; (2) for a specified exposure, common stars with good photometry (unsaturated, isolated, and magnitude error $< 0.05$ mag) were identified in each exposure and its surrounding exposures; (3) average zero point offsets of all exposures were calculated by comparing magnitudes of the common stars; (4) new offsets were iteratively obtained in the same way as above after old offsets were applied to revise the magnitudes. Finally, the internal zero point was determined by adding the final offsets to the external zero points. The number of iterations was set to be 20 and usually the offsets converged after just a few iterations. As shown in \citet{zho18}, the internal calibration could achieve a homogeneous photometric calibration over most of the sky with a precision better than 10\,mmag. We adopted the internal zero point if the number of common stars was larger than 100 and if the magnitude dispersion was less than 0.1, otherwise we used the external zero point. 

\begin{figure*}[!ht]
\centering
\includegraphics[width=0.9\textwidth]{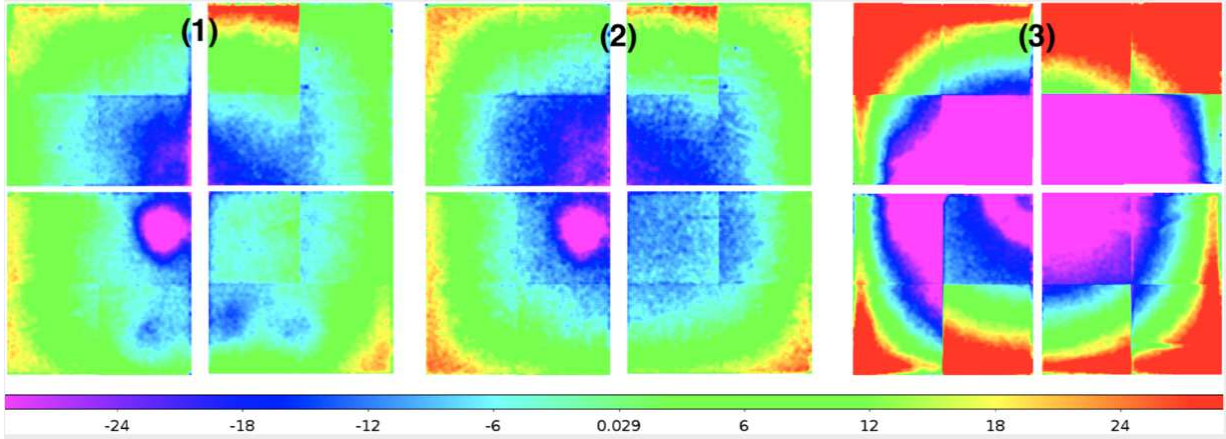}
\caption{Photometric residuals in mmag relative to the PS1 photometry. (1)--(3) photometric residuals for $g$, $r$, and $z$ bands, respectively. There are four images in each panel, representing the CCD mosaic.} \label{fig-photres} 
\end{figure*}

Photometric residuals exist across the CCD images, which are mainly caused by imperfect flat-fielding, focal distortion, the CTE effect, and scattered light. The residual maps for each band (in magnitudes) were constructed by comparing our measured magnitudes with those in the PS1 catalog using bright stars from individual images. Figure \ref{fig-photres} shows the photometric residual maps for three bands. These photometric residuals were used in the internal calibration and also used to correct the photometric measurements for the single-epoch images. 

\section{Photometric pipeline} \label{sec-photometry}
\begin{figure*}[!ht]
\centering
\includegraphics[width=\linewidth]{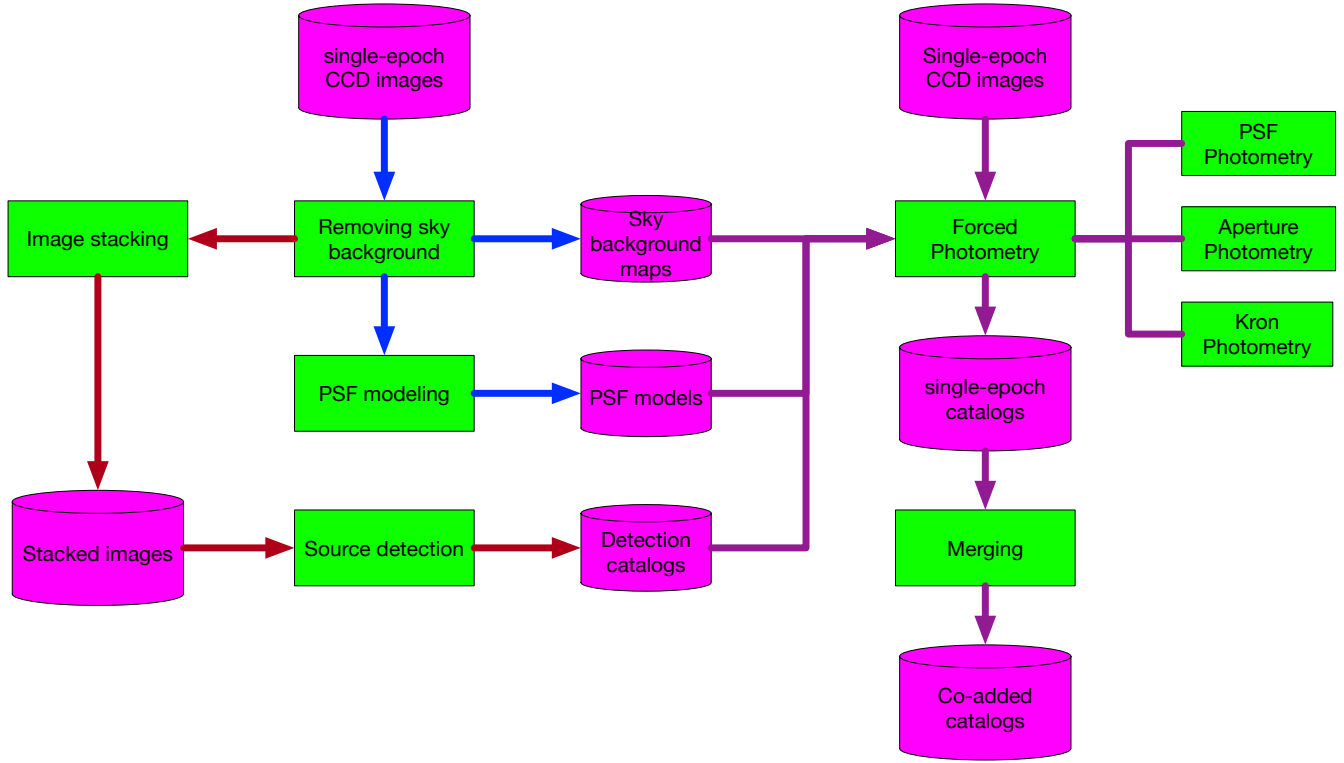}
\caption{Flowchart of the photometric pipeline.} \label{fig-flowchart} 
\end{figure*}

For most of the BASS DR3 survey area, there are at least three dithered passes, which were observed under different weather and seeing conditions. The single-epoch exposures from these passes were combined to form deeper stacked images, typically using tens of exposures taken on different nights that overlapped the area of the stack. The PSF across a stacked image typically varies dramatically, so can be very difficult to model accurately. However, the PSFs across single-epoch images typically change smoothly and can be well-modeled, meaning that single-epoch photometry can be more accurate than stacked photometry. Thus, our photometric strategy was as follows: (1) sources were detected in stacked images; (2) forced photometric measurements were made in single-epoch images with prior knowledge of the object positions from the stacks; (3) these measurements were co-added to form the final catalogs. The flowchart for the photometric pipeline is presented in Figure \ref{fig-flowchart}.  The photometric software was developed in Python\footnote{\url{https://www.python.org/}} and the stacking procedure was based on the SWarp software\footnote{\url{http://www.astromatic.net/software/swarp}}.

\subsection{Image Stacks}\label{sec-stacking}
To facilitate the stacking of images, we
divided the sky into equal-area blocks of size 0\arcdeg.68, corresponding to an image size of 5400 pixels with a pixel scale of 0\arcsec.454. Adjacent blocks have overlaps of 0\arcdeg.02 in both the right ascension and declination directions. The stacked images were then generated with a simple tangent-plane projection around the block center. We constructed stacks for each of the $g$, $r$, and $z$ bands, and an additional composite combining these three individual stacks. The stacking procedure we adopted generates both stacked images and the corresponding weight images.

The SWarp software was used to resample and co-add the single-epoch images. Median combination was used to remove outlying pixels (e.g.\ bad pixels, cosmic rays, and CCD artifacts). The sky background map for each single image was determined to avoid any overestimation caused by large extended sources or very bright stars. Specifically, a rough sky background and associated RMS map were first obtained using a mesh grid with a size of 380 pixels. The sky and RMS value in each grid point were calculated as the mode and standard deviation of the remaining pixels after outliers were removed by a $3\sigma$-clipping algorithm. Pixels containing ``signal'' from celestial objects were identified as those that were above the sky+1.5$\times$RMS. Finally, accurate sky and RMS maps were determined using the same procedure after masking the ``signal'' pixels. The area occupied by large galaxies or very bright stars was adaptively masked to cover any faint extended halos.

To ensure the stacking quality and survey homogeneity, we selected the single-epoch images that satisfied the following conditions: 

\begin{enumerate}
\item exposure time $>$ 30 s;
\item seeing $<$ 3\arcsec.5 for BASS and $<$ 2\arcsec.5 for MzLS;
\item sky ADU $<$ 15{,}000 for BASS and $<$ 25{,}000 for MzLS;
\item astrometric RMS error $<$ 0\arcsec.5;
\item number of stars used for calculating the zero point $> 50$ and zero point RMS error $<$ 0.2 mag;
\item transparency $>$ 0.5, where transparency, reflecting the atmospheric absorption, was calculated based on the zero point and airmass;
\item estimated imaging depths: $g > 21$, $r > 20.5$, and $z > 20.5$ mag.
\end{enumerate}	

\subsection{Source detection}
For each block, sources were detected in four stacked images.  
The threshold for signal detection was set to $1\sigma$ above the sky background, so that most faint objects could be identified. Sources detected in the four stacks were cross-matched with a radius of 2 pixels. We required that real objects should be detected in at least two bands, or in one band in the rare case that only one of the three bands was observed. This detection procedure was designed to remove most of the CCD artifacts, possible transient events (e.g.\ supernovae), and moving objects (e.g.\ asteroids). Because there are overlaps between blocks, objects are not unique. For a specified block, all detected objects were cross-matched with those in adjacent blocks using a matching radius of 1\arcsec.5. The duplicated sources in overlapping regions were removed according to their respective distances to the centers of their associated blocks. The object with the shortest distance to the block center was retained.

\subsection{Photometry and co-addition}
Photometry was derived using single-epoch images with object positions measured in the stacked images (stacked as outlined in Section \ref{sec-stacking}). The single-epoch photometric measurements were co-added to produce final catalogs. Before photometry was performed, objects were first projected onto individual images. The sky and associated RMS maps of an image were then estimated using the same method as described in Section \ref{sec-stacking}. Pixels with signal at $1\sigma$ above the sky background were segmented and assigned to the objects. If the pixel value of an object was less than $1\sigma$ above the sky background, we assigned a 2$\times$2 pixel-segment to the object. The object position was then redetermined based on the pixel-segments. Three types of photometric measurements were conducted, which are briefly outlined below. 

\textit{\textbf{Circular aperture photometry.}} A total of 12 apertures were used for the circular aperture photometry. The aperture diameter ranged from 6 to 80 pixels as listed in \citet{zou18}. 

\textit{\textbf{Isophotal and Kron elliptical aperture photometry.}} 
Isophotal photometry was calculated by integrating pixel fluxes within the segment of each object. At the same time, shapes including axis lengths, ellipticity, and position angle were also calculated. For better photometric measurements for galaxies, an appropriate elliptical aperture was defined via the ``Kron aperture" \citep{kro80}. The elliptical aperture is determined in a similar way as in SExtractor \citep{ber96}. For both the circular aperture and the Kron elliptical photometry, each pixel was divided into a $5\times5$ grid in order to obtain more accurate flux statistics within the apertures. If an object has masked area by nearby objects within the apertures, the flux for the masked area is complemented  by mirroring the image relative to the object center. 

\textit{\textbf{PSF photometry.}} PSF photometry is most accurate for stars and quasars. The PSF profile for a given single-epoch image was derived via a similar algorithm as adopted in PSFEx \citep{ber11}, using a linear combination of pixel basis-vectors. The spatial variation was modeled using a second-order polynomial with a PSF size of 51$\times$51 and a sampling step of 0.5\,pixels. The PSF profile was iteratively determined after removing outliers using a $3\sigma$-clipping algorithm. For objects that were not isolated (i.e., objects with segments that were connected to each other), then at most three objects were fitted simultaneously \citep{zou18}.

Each of the photometric magnitudes listed above were derived for each single-epoch image and were aperture-corrected using the growth curve derived from the circular aperture photometry. The magnitudes were scaled to the aperture magnitude with an aperture diameter of 26 pixels, which was adopted for the photometric flux calibration mentioned in Section \ref{sec-zeropoint}. The astrometric and photometric residual maps, as presented in Section \ref{sec-astrometry} and \ref{sec-zeropoint}, were used to correct the astrometric and photometric measurements in the single-epoch catalogs. 

Multiple photometric measurements from single-epoch catalogs for the same object were combined to generate deeper, co-added catalogs. We used the weighted averages (weighted by inverse variances) of these measurements, including position, shape, and flux/magnitude. For each parameter, the standard deviation of multiple measurements was also computed. The minimum, maximum and average seeing, sky brightness and number of exposures, and the observation time in Julian day, were calculated for each object. 															

\section{Quality Assessment of This Data Release} \label{sec-quality}
\subsection{Completeness and false rate of source detection}
Sources in DR3 are detected in stacked images and are required to be identified in at least two bands. This naturally removes moving objects (e.g.\ asteroids) and transients 
(e.g.\ novae and supernovae). 

We can check the completeness and false detection rate of DR3 using reference catalogs from the Canada-France-Hawaii Telescope Legacy Survey (CFHTLS\footnote{\url{http://www.cfht.hawaii.edu/Science/CFHLS/}}). More specifically, using the ``CFHTLS Deep" imaging survey. There are four independent pointings for the deep survey, yielding 80\% completeness limits of $g=26.0$, $r=25.6$, and $z=25.0$ (AB magnitudes) for point sources. These limits are about 2 magnitudes deeper than the BASS and MzLS surveys. The median seeing of the CFHTLS deep fields in $r$ band is about 0\arcsec.7. One of the four deep fields, CFHTLS-D3 ($\alpha$ = 14:19:27, $\delta$ = +52:40:56) lies within the BASS footprint. We use the photometric catalog for this field in the CFHTLS T0007 Release \citep{hud12} to estimate the completeness and false rate of source detection for BASS DR3. 

\begin{figure*}[!ht]
\centering
\includegraphics[width=\linewidth]{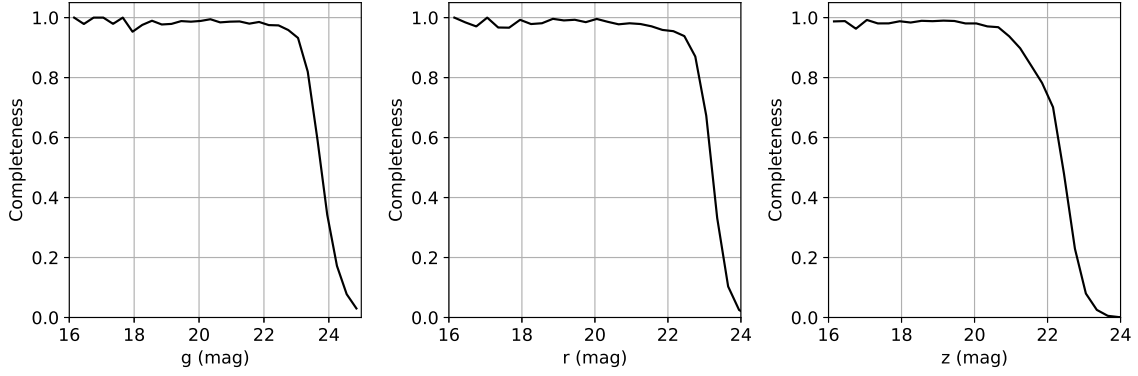}
\caption{The detection completenesses as functions of the Kron magnitudes for $g$, $r$, and $z$ bands. } \label{fig-detectcomplete} 
\end{figure*}

As shown in Figure \ref{fig-detectcomplete}, the 80\% completenesses for DR3 are 23.4, 22.9, and 21.7 mag for $g$, $r$, and $z$ bands, respectively, and the total false detection rate is about 7.4\%. By visually checking these ``false detections," we estimate that more than half of them are, in fact, real sources. Thus we expect the false detection rate to actually be less than 3\%.  The COSMOS field was also observed by BASS, but the COSMOS depths are shallower than the nominal BASS depths. As compared with the ultra-deep COSMOS catalog \citep{lea07}, the DR3 false detection rate is about 1.8\%, and, further, about one-third of these false detections appear to be real objects on visual inspection.

\subsection{Star-galaxy separation}
The magnitude difference of point-like and extended sources is adopted as a classifier in imaging surveys such as SDSS, the Hyper Suprime-Cam Subaru Strategic Program \citep[][HSC-SSP]{aih18}, and PS1. This kind of classification should perform worse at fainter magnitudes, as more distant galaxies are harder to resolve. However, additional information beyond morphology, such as colors, can improve star-galaxy classification accuracy. We therefore consider both morphology and colors when characterizing star-galaxy separation in BASS DR3.

Deeper and higher-resolution imaging surveys are needed to provide fiducial source types and assess classification accuracy. An examination with the COSMOS classification of \citet{lea07} shows that star/galaxy separation in the CFHTLS catalogs is insufficiently reliable to directly characterize BASS DR3. We therefore adopt a similar method to that applied by \citet{lea07} to separate stars from galaxies in CFHTLS-D3. The \citet{lea07} method is based on the fact that point sources occupy a well-defined locus in the MU\_MAX and MAG\_AUTO plane, because the light distribution of a point source scales with magnitude. We use the magnitude difference between Kron and PSF magnitudes as the star/galaxy classifier for each band. Such a magnitude difference is close to zero for point sources and is negative for extended sources. The thresholds are set to maximize the total success rate of both star and galaxy types when compared with the CFHTLS-D3 classification. Objects are classified as stars in each band if the magnitude differences satisfy the following criteria:

\begin{equation}
 \begin{aligned}
g_\mathrm{KRON} - g_\mathrm{PSF} > -0.101, \\
r_\mathrm{KRON} - r_\mathrm{PSF} > -0.093, \\
z_\mathrm{KRON} - z_\mathrm{PSF} > -0.047.
\end{aligned}
\label{morphonly}
\end{equation}

\begin{figure*}[!ht]
\centering
\includegraphics[width=\linewidth]{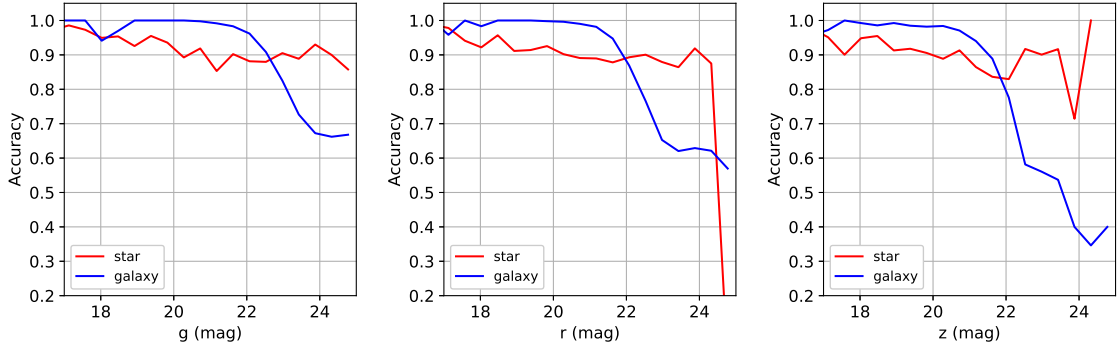}
\caption{Star-galaxy classification accuracy based on the difference of the Kron and PSF magnitudes. The classification from CFHTLS-D3 is used as a reference.} \label{fig-type} 
\end{figure*}

The global type is determined from the band-specific types in the order $z$, $r$, $g$. Note that this classification method can produce a classification even for objects that are observed in only a single band. Figure \ref{fig-type} shows the classification accuracy as a function of Kron magnitude relative to the classifications in CFHTLS-D3. The accuracy for stars (galaxies) is defined as the fraction of stars (galaxies) in our classification that are also classified as stars (galaxies) in CFHTLS-D3.

\begin{figure*}[!ht]
\centering
\includegraphics[width=\linewidth]{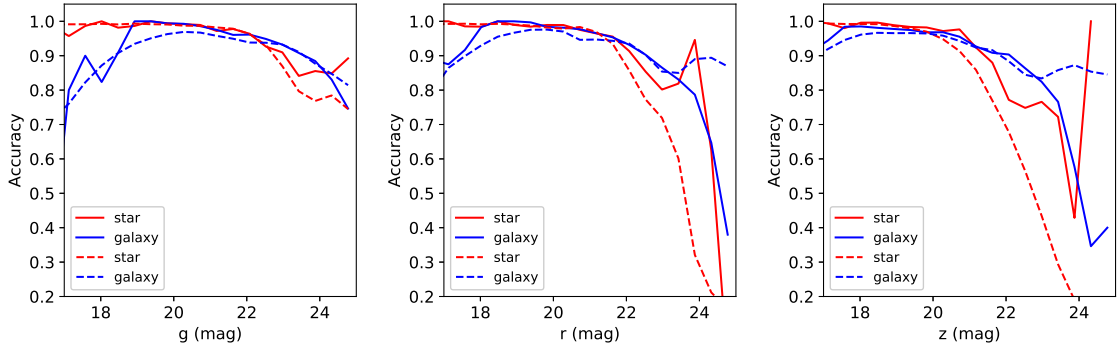}
\caption{Star-galaxy classification accuracy based on an ANN. The solid line is for the classification accuracy using the classifications from the CFHTLS-D3 deep field as the reference. The dashed line is for the classification accuracy using the combined classifications of SDSS and HSC-SSP in the HECTOMAP field as the reference.} \label{fig-typeann} 
\end{figure*}

In addition to star-galaxy classification based on morphology alone, we combine colors and morphologies to produce an alternative classification using a supervised machine learning algorithm, namely an artificial neural network (ANN). The training set for the ANN comprises the objects labelled as having reliable classifications in the CFHTLS-D3 field. Equal numbers of stars and galaxies are selected, which are randomly divided into training and testing sets. The input observables include: (1) colors of $g - r$, $r - z$, $g - z$; (2) ellipticity defined as $1-B/A$, where $A$ and $B$ are the half major and minor axes, respectively; (3) KRON\_RADIUS/FWHM\_ISO indicating the size difference between extended and point-like sources, where KRON\_RADIUS is the Kron radius and FWHM\_ISO is the FWHM of the isophotal profile. Figure \ref{fig-typeann} presents the classification accuracy for the objects in the testing set using our trained ANN. The accuracy can be as high as 80\% near the magnitude limits of DR3. The ANN method requires photometric measurements for all three bands but should be more accurate than the classification based solely on magnitude differences (Eqn.\,\ref{morphonly}). 

In order to check the field-to-field variation and reliability of the ANN classification outside the CFHTLS-D3 field, we obtain the star-galaxy separation  in the HECTOMAP field from the HSC-SSP survey \citep{aih18}. The HECTOMAP field was observed by the 8.2 m Subaru telescope, covering an area of $\sim90$ deg$^2$ in the BASS footprint. The $5\sigma$ survey depth is $r\approx26$. Due to the incompleteness and saturation of objects at the bright magnitude end, we combine the SDSS classification for objects with $r<20$. Both SDSS and HSC-SSP classifications are based on the magnitude difference between ``CModel" (a kind of galaxy model photometry) and PSF. The combined classification has an average accuracy of about 95\% at $r<24$ as estimated by the COSMOS field. Figure \ref{fig-typeann} overplots our ANN classification accuracy as function of magnitude relative to the combined classifications of SDSS and HSC-SSP. Although based on different data set, the classification accuracy in this wide field is generally consistent with that in the compact CFHTLS-D3 field, except for the faint objects. Galaxies are dominated at the faint magnitude end so that the misclassification of stars from HSC-SSP would have significant effect on the accuracy estimation. 

\subsection{Astrometric and photometric accuracy}
\begin{figure*}[!ht]
\centering
\includegraphics[width=0.8\linewidth]{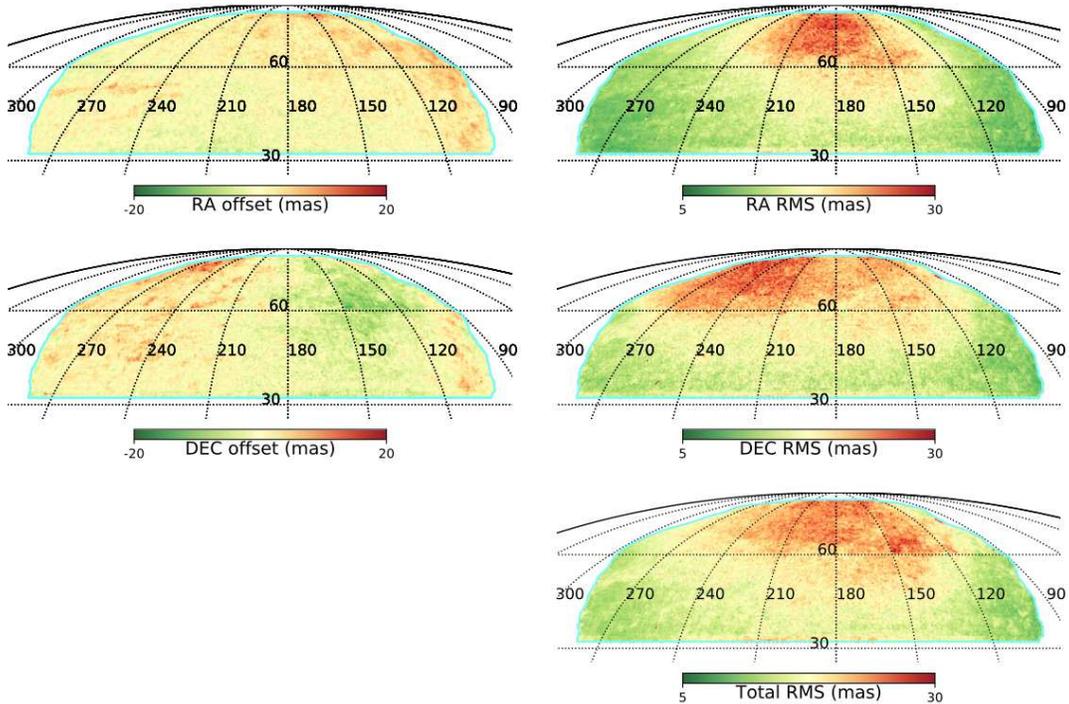}
\caption{Spatial distributions of the astrometric offsets and RMSs relative to {\it Gaia} DR2. The current DESI survey footprint is outlined in cyan.} \label{fig-astrommap} 
\end{figure*}

Figure \ref{fig-astrommap} shows the spatial distributions of the astrometric offsets and RMSs in the $g$, $r$, and $z$ bands relative to {\it Gaia} DR2. These astrometric statistics are calculated using a 3$\sigma$-clipping algorithm.  The offsets and RMSs of R.A. and decl. for 20\%, 50\% and 80\% of the survey footprint, and corresponding total astrometric precision are listed in Table \ref{tab-astrom}. In general, the median astrometric offsets relative to {\it Gaia} DR2 are less than 1\,mas and the median astrometric accuracy is about 17\,mas. The astrometric accuracy at higher declination is slightly worse than at lower declination, which is likely caused by the effect of differential chromatic refraction. 

\begin{table*}[!ht]
\centering
\caption{Astrometric offset and RMS relative to {\it Gaia} DR2}\label{tab-astrom}
 \begin{tabular}{cccccc}
 \tableline\tableline
 Percentage & R.A. offset & R.A. RMS & decl. offset & decl. RMS & Total astrometric RMS  \\
 --           &   mas       &  mas       &   mas         &  mas            & mas   \\
 \tableline
 20\%     & -1.4          &  12.2       & -3.4           &  13.9            & 14.3    \\
 50\%     & 0.2           &   14.7      & -0.6           &  16.7            & 16.9    \\
 80\%     & 2.6           &   18.5      & 1.8            &   20.3           & 20.3    \\
 \tableline
 \tableline
\end{tabular}
\end{table*}

\begin{figure*}[!ht]
\centering
\includegraphics[width=0.8\linewidth]{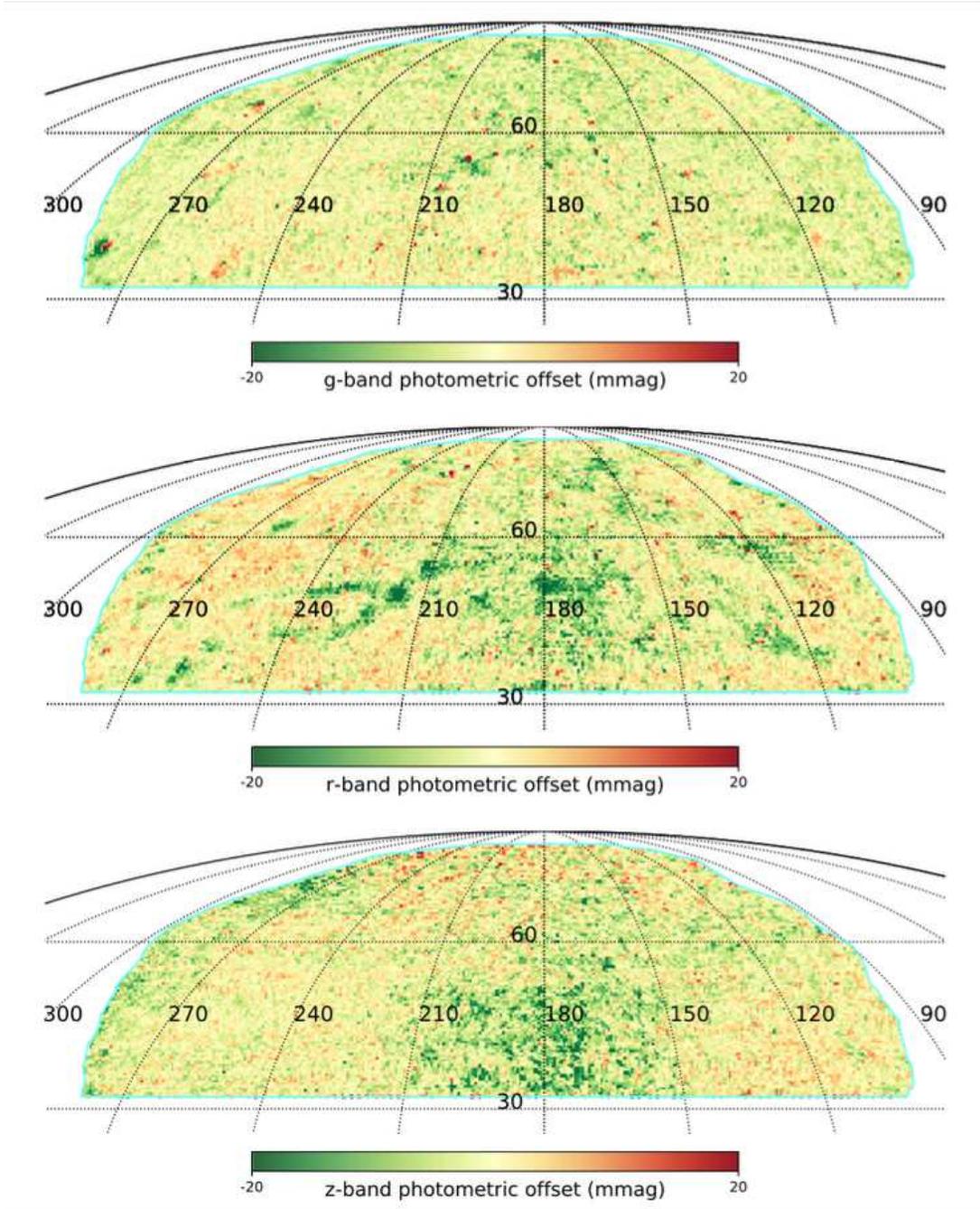}
\caption{Spatial distributions of photometric offsets relative to PS1 for $g$, $r$, and $z$ bands. The current DESI survey footprint is outlined in cyan.} \label{fig-photmap} 
\end{figure*}

To check the photometric homogeneity of BASS DR3, we compare against co-added PSF magnitudes from PS1. We use unsaturated stars with PSF magnitudes between 16 and 20.5\,mag, and calculate offsets after applying a $3\sigma$ clipping algorithm.  The PS1 magnitudes are transformed to the BASS/MzLS photometric system through the transformation equations provided in Section \ref{sec-systemequ}. Figure \ref{fig-photmap} displays the spatial distributions of the photometric offsets relative to PS1. The photometric offsets for different fractions of the survey footprint are provided in Table \ref{tab-photoffset}. The median offsets for $g$, $r$, and $z$ bands are all within 5\,mmag. The magnitude offsets for most of the survey area are $<10\,{\rm mmag}$.

\begin{table}[!ht]
\centering
\caption{Photometric offsets relative to PS1}\label{tab-photoffset}
 \begin{tabular}{cccc}
 \tableline\tableline
 Percentage & Offset ($g$) & Offset ($r$) & Offset ($z$)\\
 --           &  mmag   & mmag   & mmag     \\
 \tableline
 20\%     &   -5.2      &  -5.7      & -6.2        \\
 50\%     &  -2.2      &  -0.7      & -1.7        \\
 80\%     &   0.7      &  2.7        &  2.2        \\
 \tableline
 \tableline
\end{tabular}
\end{table}

\subsection{Photometric depths}
\begin{figure*}[!ht]
\centering
\includegraphics[width=0.8\linewidth]{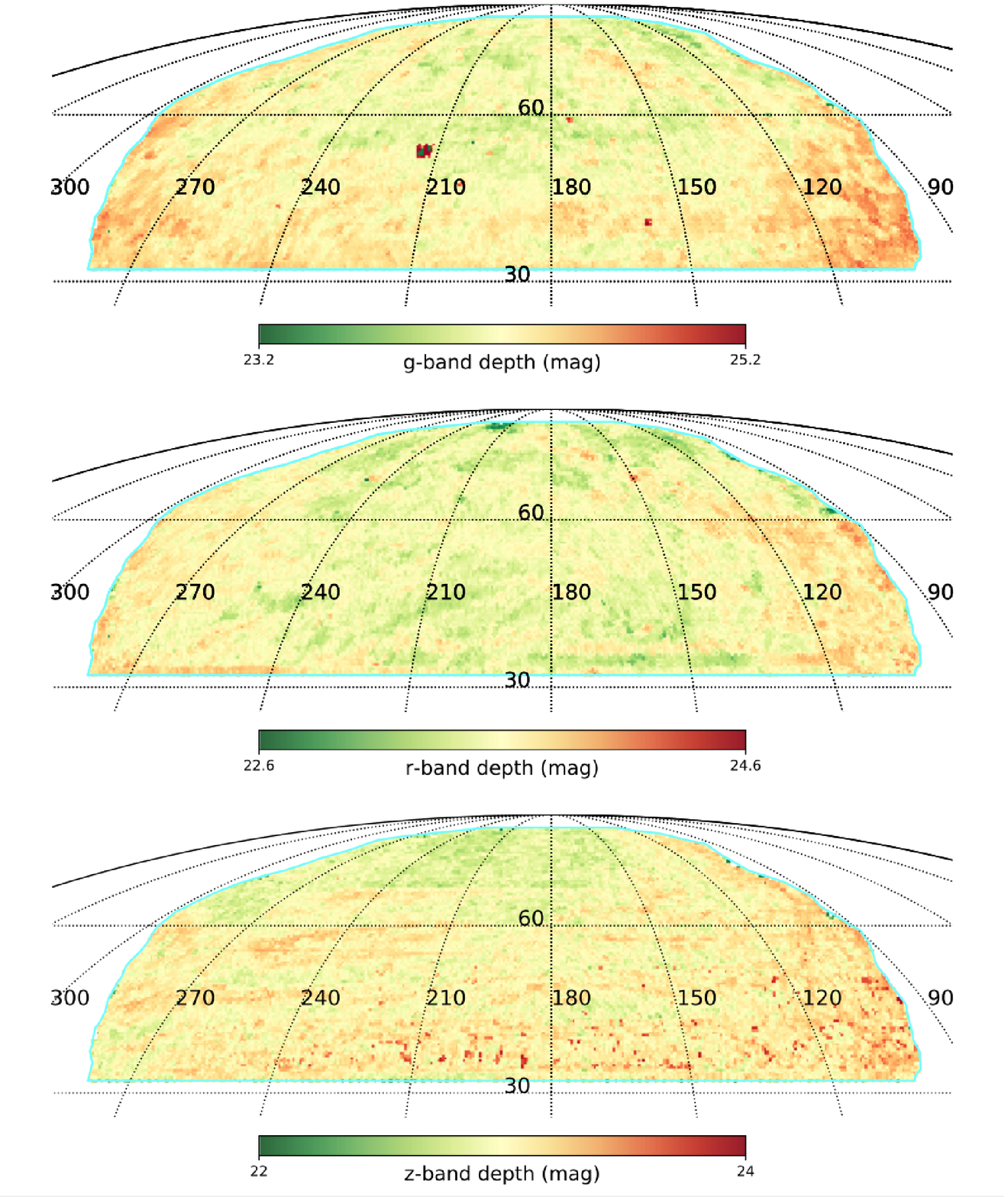}
\caption{Spatial distributions of $5\sigma$ magnitude limits for $g$, $r$, and $z$ bands. The current DESI survey footprint is outlined in cyan.} \label{fig-depthmap} 
\end{figure*}

\begin{figure}[!ht]
\centering
\includegraphics[width=\linewidth]{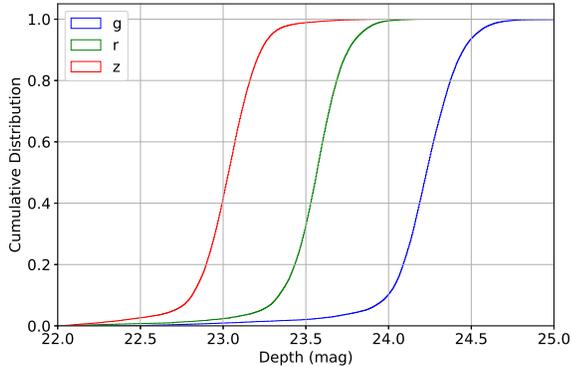}
\caption{Cumulative distributions of $5\sigma$ depths for $g$, $r$, and $z$ bands.} \label{fig-depthcdf} 
\end{figure}

We derive the $5\sigma$ magnitude limits for DR3 using the median PSF magnitude with a photometric error of $\sim0.21$ mag. Figure \ref{fig-depthmap} shows the $5\sigma$ depths for $g$, $r$, and $z$ bands. Figure \ref{fig-depthcdf} shows the cumulative distributions of the magnitude limits, and Table \ref{tab-depth} lists the depth values for different fractions of the survey area. The steep gradients of the distributions in Figure \ref{fig-depthcdf} are indicative of survey homogeneity. We derive median depths of 24.2, 23.6, and 23.0 mag for the $g$, $r$, and $z$ bands, respectively. The difference between the 20\% and 80\% depth is less than 0.3 mag.

\begin{table}[!ht]
\centering
\caption{$5\sigma$ photometric depths for PSF magnitudes}\label{tab-depth}
 \begin{tabular}{cccc}
 \tableline\tableline
 Percentage & Depth ($g$) & Depth($r$) & Depth ($z$)\\
 --           &  mag   & mag   & mag     \\
 \tableline
 20\%     &   24.09      &  23.43     & 22.89        \\
 50\%     &  24.23      &  23.56      & 23.03        \\
 80\%     &   24.37      &  23.68        &  23.16        \\
 \tableline
 \tableline
\end{tabular}
\end{table}

\begin{figure*}[!ht]
\centering
\includegraphics[width=1.0\linewidth]{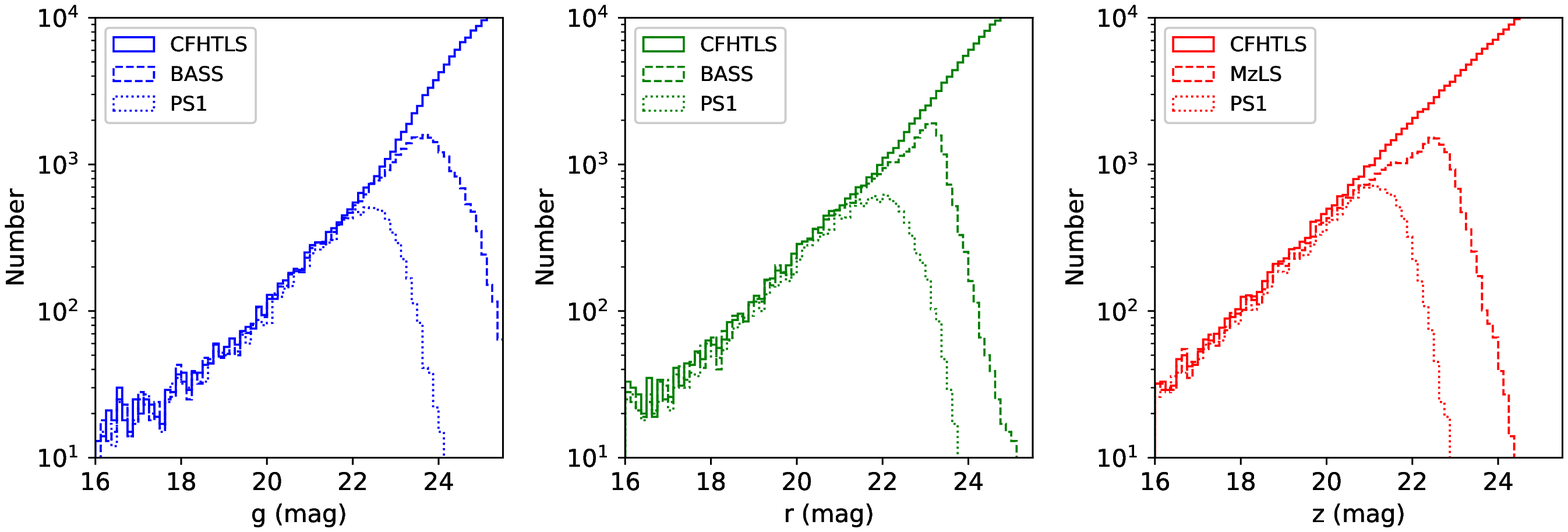}
\caption{Source number counts for BASS $g$ and $r$ and MzLS $z$ bands in the CFHTLS-D3 field (dashed lines). The number counts from CFHTLS and PS1 are plotted as solid and dotted lines, respectively.} \label{fig-sourcecount} 
\end{figure*}

Number counts of sources are also often used to estimate imaging depth. Figure \ref{fig-sourcecount} shows the number counts of BASS objects in the CFHTLS-D3 field for $g$, $r$, and $z$ bands. The peaks of the counts are located at 23.6, 23.1, and 22.4 mag for BASS $g$ and $r$ and MzLS $z$ bands, respectively. As shown in Figure \ref{fig-detectcomplete}, the corresponding completenesses are about 74\%, 78\%, and 67\%, respectively. The number counts for the CFHTLS and PS1 catalogs are also plotted for comparison. The CFHTLS catalog is regarded to be complete across the entire magnitude range of Figure \ref{fig-sourcecount}. The number counts for PS1 $g$, $r$, and $z$ bands, which peak at 22.2, 22.0, and 21.0\,mag, respectively, are more than 1\,mag shallower than BASS DR3.

\section{System transformation equations} \label{sec-systemequ}
We provide transformation equations that can be used to convert the magnitudes of other wide or deep surveys to the BASS/MzLS photometric system. The equations should also be reasonable for galaxies that do not have strong emission lines. The equations are expressed as a 3-order polynomial function of color $g-i$ (for PS1, SDSS, HSC-SSP, and CFHTLS) or $g-z$ (for DECaLS): 
\begin{equation}
m_\mathrm{BASS/MzLS} = m_\mathrm{other}+c_0+c_1X+c_2X^2+c_3X^3,
\end{equation}
where $m_\mathrm{BASS/MzLS}$ is the magnitude of a given filter of BASS $g$, $r$ or MzLS $z$, $m_\mathrm{other}$ is the corresponding magnitude of the same filter in the surveys as listed above, $X$ is the color, and $c_0$, $c_1$, $c_2$, and $c_3$ are the polynomial coefficients.  The transformation equations are derived using PSF magnitudes and are suitable for normal stars with $0.3< g-i < 2.7$ or $0.3< g-z < 4.0$. Table \ref{tab-coef} provides the coefficients from all the equations.

\begin{table}
\centering
\caption{Polynomial coefficients from the system transformation equations for different surveys.}\label{tab-coef}
 \begin{tabular}{c|cccc}
 \tableline\tableline
  Filter         & $c_0$ & $c_1$ & $c_2$ & $c_3$ \\
    \tableline
    & \multicolumn{4}{c}{PS1}       \\
    \tableline
g &  0.01310 &  0.09464 & -0.01456 & -0.00084 \\
r & -0.01068 & -0.06976 &  0.02885 & -0.01050 \\
z &  0.02376 & -0.14257 &  0.06984 & -0.01804 \\
  \tableline
    & \multicolumn{4}{c}{SDSS}      \\
  \tableline
g &  0.01759 & -0.05376 &  0.00941 & -0.00234 \\
r & -0.02841 & -0.09945 &  0.05359 & -0.01490 \\
z &  0.00540 & -0.07516 &  0.03908 & -0.00928 \\
 \tableline 
   & \multicolumn{4}{c}{HSC-SSP}   \\
    \tableline
g &  0.01852 &  0.03523 & -0.01275 &  0.00234 \\
r & -0.02199 & -0.05033 &  0.00575 & -0.00347 \\
z &  0.01652 & -0.08269 &  0.03863 & -0.00899 \\
 \tableline
   & \multicolumn{4}{c}{CFHTLS}      \\
    \tableline
g & -0.06987 &  0.09129 &  0.00267 & -0.00731 \\
r & -0.08771 & -0.06151 &  0.03295 & -0.01211 \\
z & -0.02844 & -0.10400 &  0.03805 & -0.00880 \\
 \tableline
   & \multicolumn{4}{c}{DECaLS}     \\
    \tableline
g & -0.01478 &  0.07557 & -0.01977 &  0.00232 \\
r & -0.03701 &  0.02209 & -0.00420 &  0.00074 \\
z & -0.02578 & -0.01494 &  0.00544 & -0.00103 \\
 \tableline
\end{tabular}
\end{table}


\section{Summary} \label{sec-summary} 
The BASS project is a now-complete, wide and deep imaging survey that used the 2.3\,m Bok telescope to cover an area of 5400\,deg$^2$ in the northern sky. The MzLS project covered the same area as BASS using the 4\,m Mayall telescope. In combination, these two surveys provide deep $grz$-band photometry, primarily designed to select targets for surveys with DESI. In addition to this primary goal, previous BASS imaging releases have been used for diverse scientific investigations, such as studies of high-redshift quasars, ultra-diffuse galaxies, AGNs, star-forming galaxies, strong lensing systems, and variable stars. After DESI spectroscopic follow-up, BASS/MzLS should have even broader uses ranging from studies of the Milky Way, to galaxies, large-scale structure, and the high redshift universe.  

There have been two previous releases of BASS data (DR1 and DR2). This paper summarized the third data release of BASS (DR3), which includes updated data reduction pipelines and better-characterized data. DR3 includes all BASS and MzLS data taken between January, 2015 and March, 2019. With this DR3 release, BASS achieved a median astrometric precision of 17\,mas relative to {\it Gaia} DR2 and photometric offsets of less than 5\,mmag relative to PS1. The median $5\sigma$ depths for point sources in DR3 are 24.2, 23.6, and 23.0 mag for $g$, $r$, and $z$ bands, respectively. All of the BASS data can be accessed at the official website (\url{http://batc.bao.ac.cn/BASS/doku.php?id=datarelease:home}).

\acknowledgments

This work is supported by the National Basic Research Program of China (973 Program; Grant Nos.\ 2015CB857004, 2017YFA0402600, and 2014CB845704), the National Natural Science Foundation of China (NSFC; Grant Nos.\ 11433005, 11673027, 11733007, 11320101002, 11421303, 11973038, and 11733006), and the External Cooperation Program of Chinese Academy of Sciences (Grant No.\ 114A11KYSB20160057). A.D.M.\ was supported by the Director, Office of Science, Office of High Energy Physics of the U.S. Department of Energy under Contract No.\ DE-AC02-05CH1123 and Award No.\ DE-SC0019022.

The Legacy Surveys consist of three individual and complementary projects: the Dark Energy Camera Legacy Survey (DECaLS; NOAO Proposal ID \# 2014B-0404; PIs: David Schlegel and Arjun Dey), the Beijing-Arizona Sky Survey (BASS; NOAO Proposal ID \# 2015A-0801; PIs: Zhou Xu and Xiaohui Fan), and the Mayall z-band Legacy Survey (MzLS; NOAO Proposal ID \# 2016A-0453; PI: Arjun Dey). DECaLS, BASS and MzLS together include data obtained, respectively, at the Blanco telescope, Cerro Tololo Inter-American Observatory, National Optical Astronomy Observatory (NOAO); the Bok telescope, Steward Observatory, University of Arizona; and the Mayall telescope, Kitt Peak National Observatory, NOAO. The Legacy Surveys project is honored to be permitted to conduct astronomical research on Iolkam Du'ag (Kitt Peak), a mountain with particular significance to the Tohono O'odham Nation.

BASS is a key project of the Telescope Access Program (TAP), which has been funded by the National Astronomical Observatories of China, the Chinese Academy of Sciences (the Strategic Priority Research Program ``The Emergence of Cosmological Structures" Grant No. XDB09000000), and the Special Fund for Astronomy from the Ministry of Finance. The BASS is also supported by the External Cooperation Program of Chinese Academy of Sciences (Grant No. 114A11KYSB20160057), and Chinese National Natural Science Foundation (Grant No. 11433005).

This work has made use of data from the European Space Agency (ESA) mission {\it Gaia} (\url{https://www.cosmos.esa.int/gaia}), processed by the {\it Gaia} Data Processing and Analysis Consortium (DPAC, \url{https://www.cosmos.esa.int/web/gaia/dpac/consortium}). Funding for the DPAC has been provided by national institutions, in particular the institutions participating in the {\it Gaia} Multilateral Agreement.

The Pan-STARRS1 Surveys (PS1) and the PS1 public science archive have been made possible through contributions by the Institute for Astronomy, the University of Hawaii, the Pan-STARRS Project Office, the Max-Planck Society and its participating institutes, the Max Planck Institute for Astronomy, Heidelberg and the Max Planck Institute for Extraterrestrial Physics, Garching, The Johns Hopkins University, Durham University, the University of Edinburgh, the Queen's University Belfast, the Harvard-Smithsonian Center for Astrophysics, the Las Cumbres Observatory Global Telescope Network Incorporated, the National Central University of Taiwan, the Space Telescope Science Institute, the National Aeronautics and Space Administration under Grant No. NNX08AR22G issued through the Planetary Science Division of the NASA Science Mission Directorate, the National Science Foundation Grant No. AST-1238877, the University of Maryland, Eotvos Lorand University (ELTE), the Los Alamos National Laboratory, and the Gordon and Betty Moore Foundation.

Based on observations obtained with MegaPrime/MegaCam, a joint project of CFHT and CEA/IRFU, at the Canada-France-Hawaii Telescope (CFHT) which is operated by the National Research Council (NRC) of Canada, the Institut National des Science de l'Univers of the Centre National de la Recherche Scientifique (CNRS) of France, and the University of Hawaii. This work is based in part on data products produced at Terapix available at the Canadian Astronomy Data Centre as part of the Canada-France-Hawaii Telescope Legacy Survey, a collaborative project of NRC and CNRS.

\appendix
\section{Data Access} 
The BASS data are accessible at the official website\footnote{\url{http://batc.bao.ac.cn/BASS/doku.php?id=datarelease:home}}, via the Chinese Virtual Observatory (China-VO). The DR3 page\footnote{\url{http://batc.bao.ac.cn/BASS/doku.php?id=datarelease:dr3:home}} details how to access the data. Here we include some of the most useful links that can be used to download the data:
\begin{enumerate}
   \item Searching for raw data: \url{http://explore.china-vo.org/data/bassdr3rawdata/f};
   \item Searching for files that contain single-epoch images and catalogs: \url{http://explore.china-vo.org/data/bassdr3images/f};
   \item Searching for files that contain stacked images, detections, and co-added catalogs: \url{http://explore.china-vo.org/data/bassdr3stack/f};
   \item Searching for sources in co-added catalogs: \url{http://explore.china-vo.org/data/bassdr3coadd/f};
   \item Generating {\tt wget} scripts for downloading files: \url{http://batc.bao.ac.cn/BASS/doku.php?id=datarelease:dr3:dr3_wgetbulk:};
   \item The directory tree for expert access for BASS: \url{http://batc.bao.ac.cn/BASS/doku.php?id=datarelease:dr2:dr2_wgetbulk:}; and for MzLS: \url{http://das101.china-vo.org/bass/rawdata/index.html};
   \item The directory tree for the BASS and MzLS raw data:  \url{http://das101.china-vo.org/bass/MZLS_Raw/index.html};
   \item Viewing online color images and catalogs through the sky viewer: \url{http://skyview.china-vo.org/}.
\end{enumerate}

There are also summary tables that contain information about the images and catalogs. The website contains more detailed descriptions of these files:
\begin{enumerate}
   \item {\tt bassmzls-dr3-rawinfo.fits}: information about the raw data;
   \item {\tt bassmzls-dr3-ccdinfo.fits}: information about the single-CCD images;
   \item {\tt blocks-dr3-prior.fits}: information about the stacked images  .
\end{enumerate}

The directory tree contains the following subdirectories:
\begin{enumerate}
   \item {\tt single\_image}: calibrated single-epoch CCD images;
   \item {\tt single\_cat}: catalogs of the single-epoch CCD images;
   \item {\tt stack}:  stacked images and RGB color images;
   \item {\tt coadd\_cat}: co-added catalogs;
   \item {\tt detection}: detection catalogs for the stacked images, used to conduct forced photometry in the single-epoch images;
   \item {\tt files}: summary table files.
\end{enumerate}

\section{Some instruction about the data}
The raw data comprise about 160\,k science frames and calibration frames (e.g.\ bias, dome flats, and twilight flats) from BASS and MzLS. There are about 430\,k calibrated single-epoch CCD images. Each CCD image has corresponding weight and mask images. The zero point (zp) used for transforming flux ($f$) to magnitude ($m = 2.5\log f + \mathrm{zp}$) can be found in the image header. The summary file of single CCD images provides information on the quality of observations with each CCD, such as seeing, zero point accuracy, and sky brightness. This file can be used for filtering images of poor quality; notably, images with {\tt IMQ==1} are used to generate our stacks and single-epoch catalogs.  The zero points for the stacked images in $g$, $r$, and $z$ bands are all set to be 30 and each stack has a corresponding weight image. The photometry in stacked images is not suitable for accurate photometry as discussed in Section~\ref{sec-stacking}. 

The BASS DR3 sources, which are identified in the stacked images, have corresponding detection catalogs. Catalogs for single-epoch CCD images include forced photometry based on the prior knowledge of the object positions from the detection catalogs. The co-added catalogs are combined from these single-epoch catalogs, which provide deep and accurate photometric measurements. As discussed in Section~\ref{sec-quality}, the star/galaxy separation based on the ANN (``TYPE\_ANN" in the co-added catalogs) is expected to be better than that based on magnitude differences (``TYPE") {\em if} valid photometry exists for all three bands. The Kron magnitude can reasonably describe the brightness for both extended and point-like sources, but the PSF magnitude provides a better estimate of magnitude for point sources. There are flags for objects in each band (``FLAG\_ISO") to indicate whether the objects are contaminated by bad pixels. The column ``NOBJ\_ISO" can be used to flag objects that are close to bright stars and large extended sources (i.e.\ isolated objects have NOBJ\_ISO = 1). Additional information about the images and catalogs is available on the DR3 website. 

\begin{figure*}[!ht]
\centering
\includegraphics[width=\linewidth]{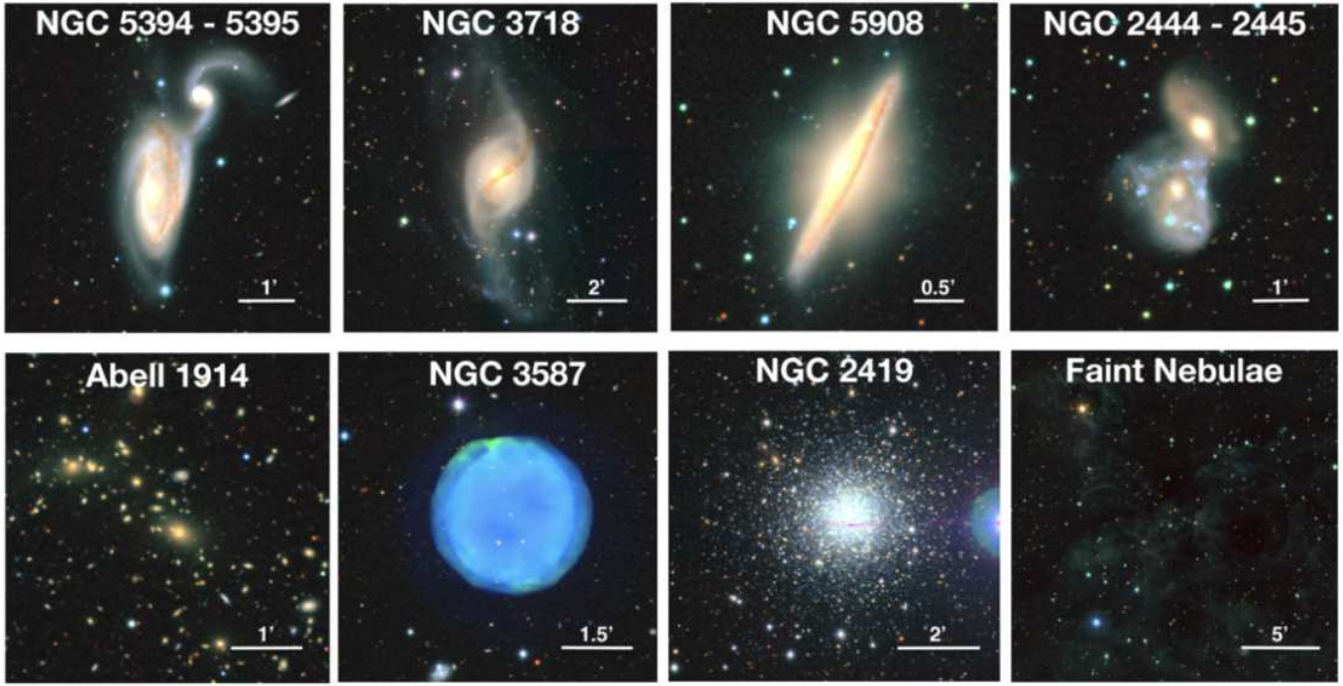}
\caption{True-color image cutouts for nearby galaxies, interacting galaxy pairs, galaxy cluster, planetary nebulae, globular cluster, and faint nebulae.} \label{fig-colorimages} 
\end{figure*}

The DR3 website includes a sky viewer that is based on Aladin HiPS\footnote{\url{http://aladin.u-strasbg.fr/hips/}}. It allows users to visualize and browse the color images and catalogs online. Figure \ref{fig-colorimages} shows several image cutouts of nearby objects.




\begin{thebibliography}{plain}
\bibliographystyle{plain}
\bibitem[Aihara et al.(2018)]{aih18} Aihara, H., Arimoto, N., Armstrong, R., et al.\ 2018, \pasj, 70, S4 
\bibitem[Bertin \& Arnouts(1996)]{ber96} Bertin, E., \& Arnouts, S.\ 1996, \aaps, 117, 393
\bibitem[Bertin(2006)]{ber06} Bertin, E.\ 2006, Astronomical Data Analysis Software and Systems XV, 351, 112 
\bibitem[Bertin(2011)]{ber11} Bertin, E.\ 2011, Astronomical Data Analysis Software and Systems XX, 442, 435 
\bibitem[Blum et al.(2016)]{blu16} Blum, R.~D., Burleigh, K., Dey, A., et al.\ 2016, American Astronomical Society Meeting Abstracts, 228, 317.01
\bibitem[Chambers et al.(2016)]{cha16} Chambers, K.~C., Magnier, E.~A., Metcalfe, N., et al.\ 2016, arXiv:1612.05560  
\bibitem[DESI Collaboration et al.(2016)]{des16} DESI Collaboration, Aghamousa, A., Aguilar, J., et al.\ 2016, arXiv:1611.00036
\bibitem[Dey et al.(2019)]{arjun19} Dey, A., Schlegel, D.~J., Lang, D., et al.\ 2019, \aj, 157, 168 
\bibitem[Gaia Collaboration et al.(2016)]{gai16a} Gaia Collaboration, Brown, A.~G.~A., Vallenari, A., et al.\ 2016, \aap, 595, A2 
\bibitem[Gaia Collaboration et al.(2016)]{gai16b} Gaia Collaboration, Prusti, T., de Bruijne, J.~H.~J., et al.\ 2016b, \aap, 595, A1 
\bibitem[Gaia Collaboration et al.(2018)]{gai18} Gaia Collaboration, Brown, A.~G.~A., Vallenari, A., et al.\ 2018, \aap, 616, A1
\bibitem[Gao et al.(2018)]{gao18} Gao, Y., Bao, M., Yuan, Q., et al.\ 2018, \apj, 869, 15 
\bibitem[Garling et al.(2018)]{gar18} Garling, C., Willman, B., Sand, D.~J., et al.\ 2018, \apj, 852, 44 
\bibitem[Hudelot et al.(2012)]{hud12} Hudelot, P., Cuillandre, J.-C., Withington, K., et al.\ 2012, VizieR Online Data Catalog, 2317
\bibitem[Kron(1980)]{kro80} Kron, R.~G.\ 1980, \apjs, 43, 305 
\bibitem[Leauthaud et al.(2007)]{lea07} Leauthaud, A., Massey, R., Kneib, J.-P., et al.\ 2007, \apjs, 172, 219 
\bibitem[Meisner, Lang \& Schlegel(2017)]{mei17} Meisner, A.M., Lang, D., \& Schlegel, D.J.\ 2017, \aj, 153, 38
\bibitem[Meisner, Lang \& Schlegel(2018)]{mei18} Meisner, A.M., Lang, D., \& Schlegel, D.J.\ 2018, RNAAS, 2, 1
\bibitem[Pickles(1998)]{pic98} Pickles, A.~J.\ 1998, \pasp, 110, 863
\bibitem[Plazas et al.(2014)]{pla14} Plazas, A.~A., Bernstein, G.~M., \& Sheldon, E.~S.\ 2014, \pasp, 126, 750   
\bibitem[Shi et al.(2017)]{shi17} Shi, D.~D., Zheng, X.~Z., Zhao, H.~B., et al.\ 2017, \apj, 846, 26 
\bibitem[Shu et al.(2018)]{shu18} Shu, Y., Marques-Chaves, R., Evans, N.~W., \& P{\'e}rez-Fournon, I.\ 2018, \mnras, 481, L136
\bibitem[Silva et al.(2016)]{sil16} Silva, D.~R., Blum, R.~D., Allen, L., et al.\ 2016, American Astronomical Society Meeting Abstracts, 228, 317.02
\bibitem[Skrutskie et al.(2006)]{skr06} Skrutskie, M.~F., Cutri, R.~M., Stiening, R., et al.\ 2006, \aj, 131, 1163  
\bibitem[The Dark Energy Survey Collaboration(2005)]{des05} The Dark Energy Survey Collaboration 2005, arXiv:astro-ph/0510346
\bibitem[Wang et al.(2017)]{wang17} Wang, F., Fan, X., Yang, J., et al.\ 2017, \apj, 839, 27
\bibitem[Wang et al.(2018)]{wang18} Wang, F., Yang, J., Fan, X., et al.\ 2018, \apjl, 869, L9 
\bibitem[Williams et al.(2004)]{wil04} Williams, G.~G., Olszewski, E., Lesser, M.~P., \& Burge, J.~H.\ 2004, \procspie, 5492, 787 
\bibitem[Wright et al.(2010)]{wri10} Wright, E.~L., Eisenhardt, P.~R.~M., Mainzer, A.~K., et al.\ 2010, \aj, 140, 1868-1881 
\bibitem[Yang et al.(2018)]{yang18} Yang, Q., Wu, X.-B., Fan, X., et al.\ 2018, \apj, 862, 109
\bibitem[Yang et al.(2019)]{yang19} Yang, J., Wang, F., Fan, X., et al.\ 2019, \aj, 157, 236 
\bibitem[York et al.(2000)]{yor00} York, D.~G., Adelman, J., Anderson, J.~E., Jr., et al.\ 2000, \aj, 120, 1579  
\bibitem[Zaritsky et al.(2019)]{zar19} Zaritsky, D., Donnerstein, R., Dey, A., et al.\ 2019, \apjs, 240, 1 
\bibitem[Zhou et al.(2018)]{zho18} Zhou, Z., Zhou, X., Zou, H., et al.\ 2018, \pasp, 130, 085001 
\bibitem[Zou et al.(2017a)]{zou17a} Zou, H., Zhang, T., Zhou, Z., et al.\ 2017, \aj, 153, 276 
\bibitem[Zou et al.(2017b)]{zou17b} Zou, H., Zhou, X., Fan, X., et al.\ 2017, \pasp, 129, 064101 
\bibitem[Zou et al.(2018)]{zou18} Zou, H., Zhang, T., Zhou, Z., et al.\ 2018, \apjs, 237, 37 
\bibitem[Zou et al.(2019)]{zou19} Zou, H., Gao, J., Zhou, X., \& Kong, X.\ 2019, \apjs, 242, 8


\end{thebibliography}
\end{document}